\documentclass[10pt]{article}
\usepackage{amsmath}
\usepackage{graphicx}
\usepackage{color}
\usepackage{orcidlink}
\usepackage{float}
\usepackage{hyperref}
\hypersetup{colorlinks=true, linkcolor=blue, citecolor=blue, urlcolor=blue}
\usepackage{caption}
\usepackage{subcaption}
\usepackage{setspace}
\usepackage{multirow}
\usepackage{multicol}
\usepackage{amssymb}
\RequirePackage[numbers,sort&compress]{natbib}
\begin{document}
\baselineskip=20pt

\begin{center}
\LARGE{Spin-($0$, $1$, $\frac{1}{2}$) Field Perturbations, Quasinormal Modes, Overtones, Greybody Factors and Strong Cosmic Censorship of Einstein-Skyrme Black Holes}
\end{center}

\vspace{0.2cm}

\begin{center}
{\bf Faizuddin Ahmed\orcidlink{0000-0003-2196-9622}}\footnote{\bf faizuddinahmed15@gmail.com}\\
{\it Department of Physics, The Assam Royal Global University, Guwahati, 781035, Assam, India}\\

{\bf  Ahmad Al-Badawi\orcidlink{0000-0002-3127-3453}}\footnote{\bf ahmadbadawi@ahu.edu.jo}\\
{\it Department of Physics, Al-Hussein Bin Talal University 71111, Ma'an, Jordan}\\

{\bf \.{I}zzet Sakall{\i}\orcidlink{0000-0001-7827-9476}}\footnote{\bf izzet.sakalli@emu.edu.tr \,(Corresponding author)}\\
{\it Physics Department, Eastern Mediterranean University, Famagusta 99628, North Cyprus via Mersin 10, Turkey}
\vspace{0.2cm}
\end{center}

\vspace{0.2cm}

\begin{abstract}
We carry out a multi-spin perturbation-theory study of the four-dimensional Einstein-Skyrme (ES) anti-de Sitter (AdS) black hole (BH), whose lapse $f(r)=1-8\pi K-2M/r+4\pi K\lambda/r^{2}$ inherits two couplings from the hadronic model -- the pion combination $K=F_{\pi}^{2}/4$ and the Skyrme coupling $e$ -- with $K\lambda=1/e^{2}$ pinned by the theory rather than being a free integration constant. After deriving the Klein-Gordon, Maxwell and Dirac effective potentials on this background, we compute the quasinormal modes (QNMs) with the sixth-order WKB formula and cross-check them against the thirteenth-order Pad\'e-improved expansion and the eikonal limit set by the unstable photon sphere. The first overtone $(n=1)$ of the scalar and electromagnetic channels reveals a mild Konoplya-Zhidenko anomaly: the ratio $|\mathrm{Im}\,\omega_{1}|/|\mathrm{Im}\,\omega_{0}|$ drifts monotonically from $2.42$ to $2.54$, sitting noticeably below the Schwarzschild value near $3$. The dominant scalar mode is independently reproduced to better than $0.2\%$ by a time-domain Prony fit. Greybody factors for all three spins follow the ordering $T_{\rm EM}<T_{\rm scalar}<T_{\rm Dirac}$. Testing strong cosmic censorship at the Cauchy horizon, we find the Christodoulou parameter $\beta\lesssim 4\times 10^{-3}$ across the admissible $(K,e)$ window -- more than two orders of magnitude below the threshold $1/2$ -- with the margin protected by the theory itself.
\end{abstract}

\tableofcontents

\section{Introduction}\label{isec1}

BHs have attracted considerable attention in recent years, particularly after the first horizon-scale images obtained by the Event Horizon Telescope (EHT) collaboration of M87$^{*}$ \cite{EHTL1,EHTL4,EHTL6} and Sgr A$^{*}$ \cite{EHTL12,EHTL16,EHTL17}. These observations have turned the near-horizon region of astrophysical BHs into a genuine laboratory for strong-field gravity, and they sit alongside a broader programme of GR tests: the predictions of Einstein's theory for the large-scale structure of the Universe remain compatible with the cosmic microwave background (CMB) measured by Planck \cite{Aghanim2020}, and its prediction of gravitational waves has been directly confirmed by LIGO and Virgo \cite{LIGO1,LIGO2,LIGO3,LIGO4}. Taken together, these successes provide independent evidence for the existence of BHs while tightening the tests of GR in the strong-field regime.

Within the standard theoretical picture, BHs are thought to form through two primary channels. The first is the gravitational collapse of massive stars at the end of their life cycles \cite{Woosley2002,Kalogera1996}: if the compact remnant exceeds the maximum mass of a neutron star, typically between two and three solar masses, it undergoes further collapse and forms a BH. The second channel is the formation of BHs in the early Universe from sufficiently large primordial density perturbations, yielding the primordial BHs (PBHs) first considered by Hawking \cite{Hawking1971}, which can span a much wider mass range than the stellar channel and remain a candidate component of dark matter.

Isolated BHs are idealized objects. In astrophysical settings they are rarely isolated; they interact with their environment and react to perturbations of the background geometry. Following such perturbations, a BH emits gravitational radiation made of a discrete spectrum of damped oscillations known as QNMs \cite{Chandrasekhar1998}. Each QNM is labelled by a complex frequency whose real part sets the oscillation frequency and whose imaginary part sets the decay rate of the perturbation, and the whole tower of modes forms the spectral fingerprint of the underlying geometry.

An important feature of QNMs is that their frequencies depend only on the intrinsic parameters of the BH and on the spin of the test field (scalar, vector, tensor or fermionic) \cite{Cavalcante2021,Konoplya2026a,Deng2026,Pani2013,Chichkov2025,Malik2025,Singh2024,Bolokhov2024,AlBadawi2023,SK2021}, and are independent of the initial perturbation that set the mode in motion. This makes QNMs sensitive probes of the underlying geometry. During the ringdown stage following a BH merger, the newly formed remnant behaves as a perturbed BH that settles down to equilibrium by emitting gravitational waves \cite{LIGO5}. The ringdown phase is dominated by QNMs, which encode the parameters of the remnant and underlie BH spectroscopy \cite{Konoplya2011,Bolokhov2025,Kokkotas1999,Konoplya2023,Konoplya2019,Berti2009}. The study of QNMs also extends beyond astrophysics and connects to BH stability, gauge/gravity duality and the quantization of horizon area. For a complete treatment of BH perturbation theory we refer the reader to the monograph by Chandrasekhar \cite{Chandrasekhar1998}.

More recently, Konoplya and Zhidenko \cite{Konoplya2024} have pointed out that \emph{overtones} of the QNM spectrum, $n\ge 1$, carry information about the near-horizon geometry that is largely invisible to the fundamental mode $n=0$. Even a small deformation of the metric localised close to the horizon can leave the $n=0$ mode essentially unchanged while shifting the first few overtones by several per cent, the so-called \emph{overtone anomaly}. The mechanism is simple enough to state: the fundamental mode is supported near the top of the effective barrier and therefore samples predominantly the curvature of the potential around its peak, while the first overtone has a longer imaginary part and its wavefunction extends deeper into the near-horizon tail of the potential, so that any short-range deformation of the geometry is amplified in the overtone channel even when it leaves the fundamental untouched. This diagnostic has since been applied to a variety of modified BHs and matter-coupled backgrounds \cite{Konoplya2024}, and it fits naturally within the ES-AdS problem where two competing couplings $(K,e)$ deform the geometry in a controlled way.

GFs describe how the pure blackbody spectrum predicted by Hawking is distorted by the effective potential barrier outside the BH horizon, which partially reflects outgoing modes. They depend on the BH geometry, on the spin of the emitted field, and on any surrounding matter. Al-Badawi and co-workers examined scalar and Dirac perturbations around NUT BHs and Schwarzschild BHs surrounded by quintessence, and showed that quintessence affects both the transmission probability and the energy flux of Hawking radiation \cite{AlBadawi2022,AlBadawi2020b,AlBadawi2020c,AlBadawi2023}. Kanzi et al. computed GFs for BHs in dRGT massive gravity and for Kerr-like BHs in Bumblebee gravity, showing that Lorentz-breaking or massive-graviton modifications of GR alter the emitted spectrum \cite{Kanzi2020,Kanzi2021,Sakalli2022c}. Further studies by Gogoi et al. and by Hosseinifar et al. have addressed greybody bounds and QNMs for BHs in Rastall gravity and in other modified gravity scenarios \cite{Gogoi2024,Hosseinifar2024}, while Sekhmani et al. analyzed the combined effect of modified Chaplygin gas and quintessence on the propagation of fields around BHs \cite{Sekhmani2025}. A topical review on GFs in various theories has been given in \cite{Sakalli2020}.

A complementary probe, rooted in the stability of GR as a classical theory, is SCC at the Cauchy horizon of BHs that possess one. In its modern (Christodoulou) formulation, SCC states that the perturbation at the Cauchy horizon should not be sufficiently regular for the field equations to be extended beyond it, i.e. not belong to $H^{1}_{\rm loc}$, which translates into the requirement
\begin{equation}
\beta \equiv \frac{|\mathrm{Im}\,\omega_{0}|}{\kappa_{-}} < \frac{1}{2},
\label{SCCcond}
\end{equation}
where $\omega_{0}$ is the dominant (slowest-decaying) QNM among all propagating spins and $\kappa_{-}$ is the surface gravity of the inner (Cauchy) horizon \cite{Cardoso2018scc,Dias2018scc,Hod2018scc,Dias2019scc}. The physics behind Eq.~\eqref{SCCcond} is that late-time perturbations decay near the Cauchy horizon as $e^{-\kappa_{-}(\beta-1/2)\,v}$ in the advanced time $v$, so that $\beta<1/2$ is precisely the borderline of the Sobolev regularity required to extend the metric across the inner horizon in a weak sense. In the Reissner-Nordstr\"om-de\,Sitter (RN-dS) family this bound is routinely violated in a narrow strip near extremality \cite{Cardoso2018scc,Dias2018scc}, giving rise to an apparent tension between the predictions of GR and the SCC conjecture. The ES-AdS BH possesses two horizons whenever $4\pi K\lambda(1-8\pi K)<M^{2}$, so SCC admits a clean and non-trivial test in the present background; the outcome, as we shall see, is structurally different from the RN-dS case.

Nonlinear field theories play a central role in many branches of physics, including quantum magnetism, the quantum Hall effect, meson theory and string theory. A well-known example is the nonlinear sigma model. Adding a Skyrme term to the nonlinear sigma model permits the construction of static soliton solutions in $3+1$ dimensions \cite{Skyrme1961a,Skyrme1961b,Skyrme1962}. Obtaining exact solutions of the Skyrme equations is difficult due to their nonlinear nature \cite{Canfora2013a}, and one therefore resorts to suitable ans\"atze such as the hedgehog ansatz. The ES system, a self-gravitating extension of the Skyrme model, was first used to build spherically symmetric BHs numerically \cite{Droz1991} and analytically via the hedgehog ansatz \cite{Canfora2013b}. Other solutions within the ES framework have been reported in \cite{Canfora2014,Canfora2018,AyonBeato2016,Astorino2018}. BHs in the ES model are important counterexamples to the no-hair conjecture, which states that BHs are fully characterized by their mass and electromagnetic charges.

The coupled ES equations admit a static, spherically symmetric AdS BH solution \cite{Skyrme1961a,Skyrme1961b,Skyrme1962,Canfora2013a,Canfora2013b,Canfora2014,Canfora2018}, described by the line element
\begin{equation}
    ds^{2} = -f(r)\,dt^{2} + \frac{dr^{2}}{f(r)} + r^{2}\,\bigl(d\theta^{2} + \sin^{2}\theta\,d\varphi^{2}\bigr),
    \label{metric}
\end{equation}
with lapse function
\begin{equation}
    f(r) = 1 - 8\pi K - \frac{2M}{r} + \frac{4\pi K\lambda}{r^{2}},
    \label{function}
\end{equation}
where $M$ is the Nucamendi-Sudarsky mass,
\begin{equation}
    K = \frac{F_{\pi}^{2}}{4},\qquad
    \lambda = \frac{4}{e^{2}\,F_{\pi}^{2}}.
    \label{couplings}
\end{equation}
The positive couplings $(F_{\pi},e)$ are phenomenologically fixed by $F_{\pi}=0.141$\,GeV and $5\le e\le 7$ \cite{Canfora2014,Adkins1983}. The limit $\lambda\to 0$ and $K\to\eta_{0}^{2}$ reduces Eq.~\eqref{metric} to the global monopole-like geometry of Barriola and Vilenkin \cite{Barriola1989}. Although the $1/r^{2}$ contribution of the Skyrme term to $f(r)$ looks formally similar to the charge contribution in the Reissner-Nordstr\"om metric, the coefficient $4\pi K\lambda$ is not an integration constant but a product of the theory couplings, and the combination $K\lambda=1/e^{2}$ is fixed by the model. This is a structural feature with far-reaching consequences: parameters that in RN can be tuned freely to bring the inner and outer horizons arbitrarily close together are, in the ES-AdS setting, pinned by the underlying hadronic physics and therefore cannot be used to drive the geometry towards extremality. The outer and inner horizons follow from $f(r_{\pm})=0$:
\begin{equation}
r_{\pm} = \frac{M\pm\sqrt{M^{2}-4\pi K\lambda(1-8\pi K)}}{1-8\pi K}.
\label{twohorizons}
\end{equation}

Motivated by these considerations, we study spin-$0$ (scalar), spin-$1$ (EM) and spin-$\tfrac{1}{2}$ (Dirac) field perturbations of the ES-AdS BH~\eqref{metric}. We analyze how the Skyrme coupling $K$ and the pion coupling $e$ shape the effective potentials, compute the fundamental QNMs and the first overtone of the scalar and EM channels, cross-check the dominant scalar mode against an independent time-domain integration, evaluate the GFs for all three spins, and finally assess the fate of SCC at the Cauchy horizon~\eqref{twohorizons}. The present work contributes to the literature in four concrete ways. First, it delivers the first multi-spin QNM census of the ES-AdS BH, covering bosonic and fermionic probes on an equal footing. Second, it applies the Konoplya-Zhidenko overtone test to a hadronic-hair background and identifies a mild but monotonic spectroscopic anomaly that is absent from the fundamental mode alone. Third, it provides an independent time-domain Prony cross-check of the dominant scalar mode that does not rely on the WKB expansion and therefore closes the numerical-accuracy question for the QNM section. Fourth, and most importantly, it performs the first SCC test on the ES-AdS BH and shows that the Christodoulou bound is respected with a wide, theory-protected margin -- a qualitatively different outcome from the near-extremal RN-dS case. The optical-appearance, shadow and Hawking emission rate of the ES-AdS BH are presented in a companion paper and are therefore not repeated here. The rest of the paper is organized as follows. Section~\ref{isec2} treats scalar perturbations, Sec.~\ref{isec3} the EM sector, Sec.~\ref{isec4} the Dirac sector, Sec.~\ref{isec5} the fundamental QNMs including the eikonal limit, Sec.~\ref{isec6} the GFs, Sec.~\ref{isec7} the overtone spectrum and the time-domain cross-check, and Sec.~\ref{isec8} the Cauchy-horizon SCC analysis. Concluding remarks are given in Sec.~\ref{isec9}. Throughout the paper we adopt geometrized units $G=c=\hbar=1$ and set $M=1$ in all numerical scans.

\section{Scalar Perturbations: Spin-0 Massless Fields}\label{isec2}

The study of scalar perturbations of BHs is a useful starting point for the analysis of wave dynamics, linear stability and QNMs in curved spacetime \cite{Chandrasekhar1998,Vishveshwara1970,Kokkotas1999,Berti2009,Ferrari1984,Konoplya2011}. A test scalar field $\Phi$ minimally coupled to gravity obeys the Klein-Gordon equation $\Box_{g}\Phi = 0$ (or $\Box_{g}\Phi = \mu^{2}\Phi$ for a massive field) on a fixed BH background. The problem reduces to a one-dimensional wave equation, and QNMs arise from the proper boundary conditions: purely ingoing at the event horizon and purely outgoing at spatial infinity, consistent with the fact that no energy can flow outwards from the horizon and no radiation can reach the system from infinity.

The massless scalar wave equation reads
\begin{equation}
    \Box_{g}\Psi = 0 \;\Longrightarrow\; \frac{1}{\sqrt{-g}}\,\partial_{\mu}\!\bigl(\sqrt{-g}\,g^{\mu\nu}\,\partial_{\nu}\Psi\bigr)=0,\qquad
    \mu,\nu=0,\dots,3,
    \label{ff1}
\end{equation}
where $g_{\mu\nu}$ is the metric tensor, $g=\det(g_{\mu\nu})$, and $g^{\mu\nu}$ its inverse. For the metric~\eqref{metric},
\begin{equation}
    g_{\mu\nu} = \mathrm{diag}\bigl(-f,\,f^{-1},\,r^{2},\,r^{2}\sin^{2}\theta\bigr),\quad
    g^{\mu\nu} = \mathrm{diag}\bigl(-1/f,\,f,\,1/r^{2},\,1/(r^{2}\sin^{2}\theta)\bigr),\quad
    g = -r^{4}\sin^{2}\theta.
    \label{ff2}
\end{equation}

In a static, spherically symmetric background we take the ansatz
\begin{equation}
    \Psi(t,r,\theta,\varphi) = e^{-i\omega t}\,Y_{\ell}^{m}(\theta,\varphi)\,\frac{\psi(r)}{r},
    \label{ff3}
\end{equation}
with $\omega$ the (possibly complex) frequency, $\psi(r)$ the radial function, and $Y_{\ell}^{m}$ the spherical harmonics. Substituting \eqref{ff3} into \eqref{ff1} and using the tortoise coordinate defined below leads to
\begin{equation}
    \frac{d^{2}\psi(r_{\ast})}{dr_{\ast}^{2}} + \bigl(\omega^{2} - V_{\rm scalar}\bigr)\,\psi(r_{\ast}) = 0,
    \label{ff4}
\end{equation}
where we used the identity
\begin{equation}
    \left[\frac{1}{\sin\theta}\frac{\partial}{\partial\theta}\!\left(\sin\theta\frac{\partial}{\partial\theta}\right)
        + \frac{1}{\sin^{2}\theta}\frac{\partial^{2}}{\partial\varphi^{2}}\right]Y_{\ell m}(\theta,\varphi)
    = -\ell(\ell+1)\,Y_{\ell m}(\theta,\varphi),
    \label{ff5}
\end{equation}
and introduced the tortoise coordinate
\begin{equation}
    r_{\ast} = \int\frac{dr}{f(r)},\qquad \partial_{r_{\ast}} = f\,\partial_{r}.
    \label{ff6}
\end{equation}
The integral in \eqref{ff6} cannot be expressed in closed form for the ES-AdS geometry because the numerator $f(r)$ is a cubic polynomial in $1/r$ whose roots do not admit rational expression in the couplings. In the relevant range, $r_{\ast}\to -\infty$ as $r\to r_{+}$ and $r_{\ast}\to +\infty$ as $r\to\infty$.

The scalar effective potential reads
\begin{equation}
    V_{\rm scalar}(r) = \left(\frac{\ell(\ell+1)}{r^{2}} + \frac{f'(r)}{r}\right)f(r)
    = \frac{1}{r^{2}}\!\left(\ell(\ell+1) + \frac{2M}{r} - \frac{8\pi K\lambda}{r^{2}}\right)
      \left(1 - 8\pi K - \frac{2M}{r} + \frac{4\pi K\lambda}{r^{2}}\right),\quad \ell\ge 0.
    \label{ff7}
\end{equation}
The second form makes two features manifest. The leading $\ell(\ell+1)/r^{2}$ piece is the usual centrifugal barrier, while the $f'(r)/r$ correction produces a mass-like and a Skyrme-charge-like contribution with opposite signs; the $-8\pi K\lambda/r^{4}$ short-range term is therefore genuinely repulsive and is responsible for the attractive character of the Skyrme hair at small $r$ -- a feature that will later control the location of the Cauchy horizon.

\begin{figure}[ht!]
    \centering
    \includegraphics[width=0.45\linewidth]{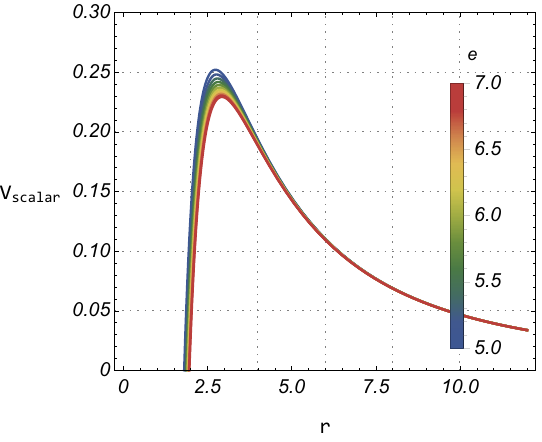}\qquad
    \includegraphics[width=0.45\linewidth]{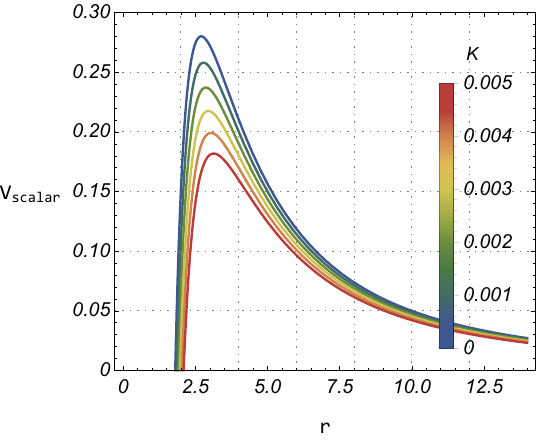}\\
    (i) $K=0.002$ \hspace{6cm} (ii) $e=6$
    \caption{\footnotesize Behaviour of the scalar effective potential $V_{\rm scalar}$ as a function of the radial distance $r$, for $M=1$, $\ell=2$. Left: $K=0.002$ fixed, $e$ varying. Right: $e=6$ fixed, $K$ varying.}
    \label{fig:effPot}
\end{figure}

Figure~\ref{fig:effPot} displays the scalar effective potential $V_{\rm scalar}$ for the $\ell=2$ multipole under variations of the pion coupling $e$ (left panel) and of the Skyrme coupling $K$ (right panel). Each curve exhibits the canonical shape of a single-peak barrier outside the event horizon, with $V_{\rm scalar}\to 0$ at the horizon and a slow decay at large $r$ set by the centrifugal term $\ell(\ell+1)/r^{2}$. The monotonic decrease of the barrier at large $r$, together with its regular behaviour near the horizon, guarantees that the QNM eigenvalue problem is well posed and that no long-lived trapped modes can form at the top of the barrier. In the left panel the family of curves lies very close to one another: raising $e$ from $5$ to $7$ produces only a mild downward displacement of the maximum, signalling the subleading role played by the pion coupling in shaping the near-horizon geometry. In the right panel, by contrast, the maximum slides upward appreciably when $K$ grows from $0$ to $0.005$, and the location of the peak shifts outward, reflecting the dual effect of $K$: it rescales the asymptotic value of $f(r)$ through the $1-8\pi K$ factor and strengthens the $1/r^{2}$ contribution. These two panels already suggest that the $K$ dependence dominates the scalar QNM response discussed in Sec.~\ref{isec5}, while $e$ acts as a fine tuning on top of it.

\section{EM Perturbations: Spin-1 Fields}\label{isec3}

Electromagnetic (spin-1) perturbations describe small disturbances of the electromagnetic field propagating on a fixed background spacetime. They are governed by Maxwell's equations and are widely used to study wave dynamics and stability properties of BH systems, where the massless, conformally invariant nature of the electromagnetic field turns the barrier into a purely centrifugal one. On the ES-AdS background this will translate into a cleaner, more transparent comparison with the scalar and Dirac sectors.

We now consider EM perturbations on the ES-AdS background. In curved spacetime, the free EM field obeys the source-free Maxwell equations \cite{Zhang2020}
\begin{equation}
    \frac{1}{\sqrt{-g}}\,\partial_{\mu}\!\bigl(\sqrt{-g}\,g^{\mu\sigma}\,g^{\nu\tau}\,F_{\sigma\tau}\bigr) = 0,
    \label{em1}
\end{equation}
with $F_{\sigma\tau}=\partial_{\sigma}A_{\tau}-\partial_{\tau}A_{\sigma}$ and $A_{\mu}$ the vector potential. In a spherically symmetric background, we employ the Regge-Wheeler-Zerilli (RWZ) formalism and decompose $A_{\mu}$ into scalar and vector spherical harmonics:
\begin{equation}
    A_{\mu} = \sum_{\ell,m} e^{-i\omega t}\!\left[
    \begin{pmatrix}0\\[1pt] 0\\[1pt] \psi_{\rm em}(r)\,{\bf S}_{\ell,m}\end{pmatrix}
    + \begin{pmatrix}j^{\ell,m}(r)\,Y_{\ell}^{m}\\[1pt]
                     h^{\ell,m}(r)\,Y_{\ell}^{m}\\[1pt]
                     k^{\ell,m}(r)\,{\bf Y}_{\ell}^{m}\end{pmatrix}\right],
    \label{em2}
\end{equation}
with the vector harmonics
\begin{equation}
    {\bf S}_{\ell,m} = \begin{pmatrix}\dfrac{1}{\sin\theta}\,\partial_{\varphi}Y_{\ell}^{m}\\[6pt] -\sin\theta\,\partial_{\theta}Y_{\ell}^{m}\end{pmatrix},\qquad
    {\bf Y}_{\ell,m} = \begin{pmatrix}\partial_{\theta}Y_{\ell}^{m}\\[4pt] \partial_{\varphi}Y_{\ell}^{m}\end{pmatrix}.
    \label{em3}
\end{equation}
The first term in Eq.~\eqref{em2} corresponds to axial (odd-parity) perturbations with parity $(-1)^{\ell+1}$, while the second represents polar (even-parity) perturbations with parity $(-1)^{\ell}$. A standard result of BH perturbation theory is that axial and polar modes give identical physical observables \cite{ref70,ref71}; this \emph{isospectrality} reflects an underlying Darboux transformation between the two sectors and holds for any static spherically symmetric background, so that we may work with the axial sector without loss of generality. Substituting the axial decomposition into Eq.~\eqref{em1} and moving to the tortoise coordinate gives the Schr\"odinger-like equation
\begin{equation}
    \frac{d^{2}\psi_{\rm em}(r_{\ast})}{dr_{\ast}^{2}} + \bigl(\omega^{2} - V_{\rm em}\bigr)\,\psi_{\rm em}(r_{\ast}) = 0,
    \label{em4}
\end{equation}
with the EM effective potential
\begin{equation}
    V_{\rm em}^{\ell}(r) = \frac{\ell(\ell+1)}{r^{2}}\,f(r)
    = \frac{\ell(\ell+1)}{r^{2}}\left(1 - 8\pi K - \frac{2M}{r} + \frac{4\pi K\lambda}{r^{2}}\right).
    \label{em5}
\end{equation}
Compared with Eq.~\eqref{ff7}, the EM potential lacks the $f'(r)/r$ term present in the scalar case. This difference stems from the vector nature of the EM field and its coupling to spacetime curvature, and it makes the EM potential insensitive to the sign of $f'(r)$. A direct consequence is that, at fixed $(\ell,M,K,e)$, the EM peak is higher than the scalar one and lies slightly further out, which already anticipates the qualitative hierarchy of GFs reported in Sec.~\ref{isec6}. The dependence on $K$, $e$ and $\ell$ remains qualitatively the same as in the scalar sector.

\section{Dirac Perturbations: Spin-1/2 Massless Fields}\label{isec4}

Massless Dirac field perturbations describe how spin-$\tfrac{1}{2}$ fermionic fields evolve under small disturbances in a given background spacetime or field configuration. They are governed by the massless Dirac equation in curved space-time, and they are a standard tool to analyze the stability and the quantum response of BH backgrounds to fermionic probes, playing a role conceptually parallel to the bosonic sectors examined in Secs.~\ref{isec2} and \ref{isec3}.

To study spin-$\tfrac{1}{2}$ perturbations we consider the Dirac equation in curved spacetime,
\begin{equation}
    \gamma^{\mu}\bigl(\partial_{\mu} + \Gamma_{\mu}\bigr)\Psi = 0,
    \label{dir1}
\end{equation}
where $\gamma^{\mu}$ are the curved-spacetime gamma matrices and $\Gamma_{\mu}$ the spin connection \cite{Unruh1973,Chandrasekhar1976}. Separating variables in terms of spinor spherical harmonics on the static spherically symmetric metric~\eqref{metric} reduces Eq.~\eqref{dir1} to a pair of decoupled Schr\"odinger-like equations,
\begin{equation}
    \frac{d^{2}F}{dr_{\ast}^{2}} + \bigl(\omega^{2} - V_{+}(r)\bigr)F = 0,\qquad
    \frac{d^{2}G}{dr_{\ast}^{2}} + \bigl(\omega^{2} - V_{-}(r)\bigr)G = 0,
    \label{dir2}
\end{equation}
where the Dirac effective potentials are
\begin{equation}
    V_{\pm}(r) = W(r)^{2} \pm \frac{dW(r)}{dr_{\ast}},\qquad
    W(r) = \frac{\kappa\,\sqrt{f(r)}}{r},
    \label{dfp1}
\end{equation}
and $\kappa$ is related to the total angular momentum by
\begin{equation}
    \kappa = \pm\!\left(j+\tfrac{1}{2}\right),\qquad j=\tfrac{1}{2},\tfrac{3}{2},\tfrac{5}{2},\dots
    \label{dir3}
\end{equation}
For the ES-AdS metric the superpotential reads
\begin{equation}
    W(r) = \frac{\kappa}{r}\sqrt{1 - 8\pi K - \frac{2M}{r} + \frac{4\pi K\lambda}{r^{2}}}.
    \label{dir4}
\end{equation}
The pair $V_{\pm}$ is supersymmetric in the sense of Darboux: the two potentials are related by the intertwining operators $\partial_{r_{\ast}}\pm W$ and therefore share the same transmission and reflection amplitudes, hence the same QNM spectrum \cite{Chandrasekhar1998}. This supersymmetry is a feature of the massless Dirac problem on any static spherically symmetric background, and it does not rely on any special property of the ES lapse; we therefore work with $V_{+}$ throughout the numerical analysis.

\begin{figure}[ht!]
    \centering
    \includegraphics[width=0.45\linewidth]{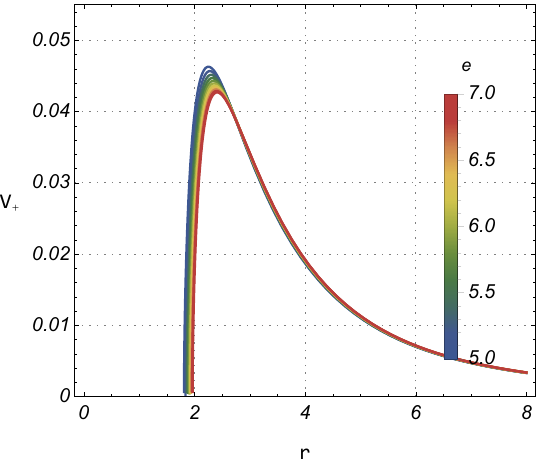}\qquad
    \includegraphics[width=0.45\linewidth]{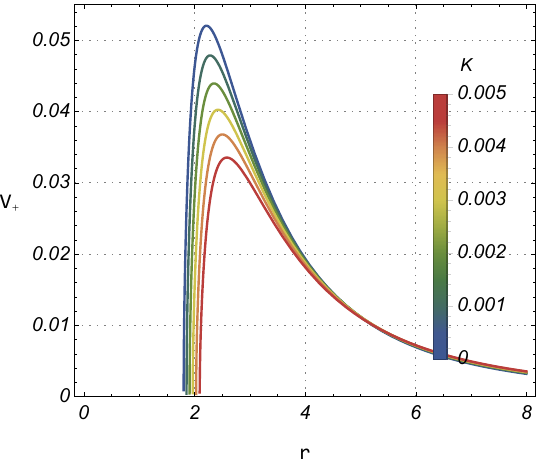}\\
    (i) $K=0.002$ \hspace{6cm} (ii) $e=6$
    \caption{\footnotesize Behaviour of the Dirac effective potential $V_{+}$ as a function of the radial distance $r$, for $M=1$, $j=1/2$. Left: $K=0.002$ fixed, $e$ varying. Right: $e=6$ fixed, $K$ varying.}
    \label{fig:effPotD}
\end{figure}

Figure~\ref{fig:effPotD} shows the Dirac potential $V_{+}$ for the lowest half-integer mode $j=1/2$ under the same parameter scans as in Fig.~\ref{fig:effPot}. The barrier is considerably lower than in the scalar case because the centrifugal contribution scales as $\kappa^{2}/r^{2}$ with $\kappa=1$ at $j=1/2$, roughly a factor of six below the $\ell(\ell+1)=6$ of the scalar reference curve. This is the direct counterpart of the well-known softening of the Dirac barrier in the Schwarzschild limit, and it provides the ES-AdS problem with a natural low-barrier channel in which subtle features of the near-horizon geometry become enhanced. In the left panel the peak position is almost insensitive to $e$: the five curves essentially collapse onto one another, with only a barely visible downward drift of the maximum as $e$ increases. The right panel shows a pronounced response to $K$: the peak rises and moves slightly outward as $K$ grows from $0$ to $0.005$, tracing the same trend observed in the scalar sector. The fact that $K$ controls the barrier height for both bosonic and fermionic fields, while $e$ merely dresses it, foreshadows the coherent multi-spin response in the QNM and GF sectors that follow.

\section{Fundamental Quasinormal Modes}\label{isec5}

QNMs characterize the response of a BH to external perturbations and supply useful information on its stability and its observational signatures, especially in the context of gravitational-wave astronomy \cite{Kokkotas1999,Konoplya2011,Berti2009}. Their spectrum depends on the intrinsic parameters of the BH and on the properties of any surrounding matter, which makes QNMs a sensitive tool for probing departures from GR and exploring the imprint of alternative gravitational models. In the present context, QNMs translate the hadronic input $(F_{\pi},e)$ of the Skyrme model into a set of complex frequencies that, at least in principle, are within reach of next-generation gravitational-wave detectors.

\vspace{0.2cm}
\begin{center}
    \large{\bf A.\,Wentzel-Kramers-Brillouin (WKB) approximation}
\end{center}

We employ the semi-analytical WKB approximation, which matches the WKB expansions near the BH horizon and at spatial infinity with a Taylor expansion of the effective potential near its peak. The WKB method was first proposed in \cite{SS2,SS3,SS4,SS5} and later extended to higher orders, including sixth \cite{SS6,SS7,SS8} and thirteenth order \cite{SS9,SS10}. Increasing the WKB order does not necessarily improve the accuracy, because the asymptotic WKB series is divergent and its optimal truncation depends on the shape of the barrier; we therefore adopt the sixth-order version and cross-check it with the thirteenth-order one, following the standard practice in recent BH spectroscopy. The sixth-order quantization condition reads
\begin{equation}
    i\,\frac{\omega_{n}^{2}-\mathcal{V}_{0}}{\sqrt{-2\,\mathcal{V}_{0}''}}+\sum_{i=2}^{6}\Phi_{i} = n+\frac{1}{2},
    \label{qnm}
\end{equation}
where $\mathcal{V}_{0}$ is the peak of the effective potential, $\mathcal{V}_{0}''$ is its second derivative with respect to the tortoise coordinate, the $\Phi_{i}$ are higher-order corrections explicitly given in \cite{SS7}, and $n=0,1,2,\dots$ is the overtone number.

The fundamental ($n=0$) QNM spectra collected in Tables~\ref{tab:qnm-scalar}--\ref{tab:qnm-dirac} show clear, monotonic trends in both $e$ and $K$. For scalar perturbations, increasing $e$ slightly reduces the real part $\mathrm{Re}(\omega)$ for every multipole number, and slightly reduces the magnitude of the imaginary part $|\mathrm{Im}(\omega)|$, indicating longer-lived modes. Raising $\ell$ enlarges the real part while leaving the imaginary part nearly unchanged, as expected from the centrifugal term, which shifts the peak of the barrier upwards without altering its curvature much. A very similar trend is found for EM modes, whose damping is slightly smaller than in the scalar sector, and for the Dirac modes. The effect of $K$ is stronger and more uniform: both $\mathrm{Re}(\omega)$ and $|\mathrm{Im}(\omega)|$ decrease with growing $K$ across all three spins, showing that $K$ effectively softens the near-horizon geometry, lowers the height and curvature of the effective potential, and generates slower, longer-lived oscillations. All computed modes have $\mathrm{Im}(\omega)<0$, which indicates stability of the ES-AdS BH under scalar, EM and Dirac perturbations and rules out exponentially growing branches of the spectrum for every value of $(K,e)$ in the phenomenological window.

\begin{table}[ht!]
\centering
\renewcommand{\arraystretch}{1.35}
\begin{tabular}{c|c|c|c}
\hline
\multicolumn{4}{c}{Scalar perturbations, sixth-order WKB}\\
\hline
$e$ ($K=0.002$) & $\ell=0$ & $\ell=1$ & $\ell=2$\\
\hline
$5.0$ & $0.097797 - 0.095531\,i$ & $0.292534 - 0.079441\,i$ & $0.489707 - 0.084323\,i$\\
$5.5$ & $0.094636 - 0.095072\,i$ & $0.286655 - 0.078250\,i$ & $0.480572 - 0.083723\,i$\\
$6.0$ & $0.092352 - 0.094718\,i$ & $0.282495 - 0.077361\,i$ & $0.474118 - 0.083240\,i$\\
$6.5$ & $0.090658 - 0.094449\,i$ & $0.279424 - 0.076680\,i$ & $0.469354 - 0.082855\,i$\\
$7.0$ & $0.089367 - 0.094241\,i$ & $0.277081 - 0.076150\,i$ & $0.465721 - 0.082545\,i$\\
\hline
$K$ ($e=6$) & \multicolumn{3}{c}{}\\
\hline
$0.001$ & $0.097633 - 0.099849\,i$ & $0.294683 - 0.081477\,i$ & $0.494221 - 0.087662\,i$\\
$0.002$ & $0.092352 - 0.094718\,i$ & $0.282495 - 0.077361\,i$ & $0.474118 - 0.083240\,i$\\
$0.003$ & $0.087237 - 0.089725\,i$ & $0.270532 - 0.073357\,i$ & $0.454359 - 0.078932\,i$\\
$0.004$ & $0.082284 - 0.084870\,i$ & $0.258792 - 0.069465\,i$ & $0.434946 - 0.074739\,i$\\
$0.005$ & $0.077493 - 0.080153\,i$ & $0.247275 - 0.065685\,i$ & $0.415880 - 0.070661\,i$\\
\hline
\end{tabular}
\caption{\footnotesize Fundamental ($n=0$) scalar QNMs of the ES-AdS BH with $M=1$, computed with the sixth-order WKB formula.}
\label{tab:qnm-scalar}
\end{table}

Table~\ref{tab:qnm-scalar} quantifies the fundamental scalar QNM spectrum of the ES-AdS BH for $n=0$ and $\ell\in\{0,1,2\}$ over the phenomenological window $5\le e\le 7$ at fixed $K=0.002$ and over $0.001\le K\le 0.005$ at fixed $e=6$. In the upper block, raising $e$ from $5$ to $7$ reduces $\mathrm{Re}(\omega)_{\ell=2}$ from $0.4897$ to $0.4657$, a shift of about $5\%$, while $|\mathrm{Im}(\omega)|$ drops by less than $3\%$, confirming that the pion coupling acts as a gentle softener of the scalar channel. In the lower block the response to $K$ is markedly stronger: $\mathrm{Re}(\omega)_{\ell=2}$ falls from $0.4942$ at $K=0.001$ to $0.4159$ at $K=0.005$, a $16\%$ reduction, and $|\mathrm{Im}(\omega)|$ drops by a comparable amount. The centrifugal hierarchy $\mathrm{Re}(\omega)_{\ell=0}<\mathrm{Re}(\omega)_{\ell=1}<\mathrm{Re}(\omega)_{\ell=2}$ is preserved throughout, and every entry carries a negative imaginary part, ruling out exponentially growing scalar modes in the explored parameter region. At fixed $\ell=2$ the quality factor $Q\equiv\mathrm{Re}(\omega)/(2|\mathrm{Im}(\omega)|)$ ranges from $Q\simeq 2.82$ at $K=0.001$ to $Q\simeq 2.94$ at $K=0.005$, so the ES-AdS ringdown is slightly more coherent than the Schwarzschild reference value $Q\simeq 2.51$ at the same multipole, and the quality factor grows monotonically with the Skyrme hair -- a trend that mirrors the softening of the potential barrier.

\begin{table}[ht!]
\centering
\renewcommand{\arraystretch}{1.35}
\begin{tabular}{c|c|c|c}
\hline
\multicolumn{4}{c}{EM perturbations, sixth-order WKB}\\
\hline
$e$ ($K=0.002$) & $\ell=0$ & $\ell=1$ & $\ell=2$\\
\hline
$5.0$ & $0.250100 - 0.069877\,i$ & $0.466150 - 0.082240\,i$ & $0.668666 - 0.085345\,i$\\
$5.5$ & $0.244062 - 0.068014\,i$ & $0.457051 - 0.081537\,i$ & $0.656063 - 0.084921\,i$\\
$6.0$ & $0.239822 - 0.066658\,i$ & $0.450636 - 0.080983\,i$ & $0.647170 - 0.084558\,i$\\
$6.5$ & $0.236707 - 0.065639\,i$ & $0.445909 - 0.080547\,i$ & $0.640612 - 0.084259\,i$\\
$7.0$ & $0.234340 - 0.064853\,i$ & $0.442308 - 0.080199\,i$ & $0.635615 - 0.084014\,i$\\
\hline
$K$ ($e=6$) & \multicolumn{3}{c}{}\\
\hline
$0.001$ & $0.249017 - 0.069822\,i$ & $0.469141 - 0.085211\,i$ & $0.674024 - 0.089048\,i$\\
$0.002$ & $0.239822 - 0.066658\,i$ & $0.450636 - 0.080983\,i$ & $0.647170 - 0.084558\,i$\\
$0.003$ & $0.230725 - 0.063553\,i$ & $0.432413 - 0.076860\,i$ & $0.620741 - 0.080183\,i$\\
$0.004$ & $0.221728 - 0.060510\,i$ & $0.414473 - 0.072841\,i$ & $0.594740 - 0.075922\,i$\\
$0.005$ & $0.212833 - 0.057527\,i$ & $0.396818 - 0.068927\,i$ & $0.569169 - 0.071777\,i$\\
\hline
\end{tabular}
\caption{\footnotesize Fundamental ($n=0$) EM QNMs of the ES-AdS BH with $M=1$, computed with the sixth-order WKB formula.}
\label{tab:qnm-em}
\end{table}

Table~\ref{tab:qnm-em} lists the fundamental EM QNMs with the same parameter scan. For every multipole, $\mathrm{Re}(\omega)$ of the EM mode exceeds the scalar counterpart at the same $(\ell,e,K)$: at $\ell=2,K=0.002,e=6$ the EM real part reaches $0.6472$ against the scalar value $0.4741$, reflecting the absence of the $f'(r)/r$ contribution in $V_{\rm em}$ which leaves a pure centrifugal barrier. The $|\mathrm{Im}(\omega)|$ values are close to those of the scalar sector but slightly smaller, so that EM modes are marginally longer-lived, consistent with their somewhat lower barrier curvature. Under variations of $K$ the decrease is even more marked than in the scalar case: $\mathrm{Re}(\omega)_{\ell=2}$ drops from $0.6740$ at $K=0.001$ to $0.5692$ at $K=0.005$, a fall of roughly $16\%$, and $|\mathrm{Im}(\omega)|$ shrinks by about $19\%$. These trends are faithfully reproduced for $\ell=0$ and $\ell=1$, confirming that the EM sector is just as sensitive to the Skyrme hair as the scalar one, but shifted to higher oscillation frequencies; in practice, the EM channel offers the best signal-to-noise ratio for constraining $K$ in a hypothetical electromagnetic-wave based ringdown detection.

\begin{table}[ht!]
\centering
\renewcommand{\arraystretch}{1.35}
\begin{tabular}{c|c|c|c}
\hline
\multicolumn{4}{c}{Dirac perturbations, sixth-order WKB}\\
\hline
$e$ ($K=0.002$) & $j=1/2$ & $j=3/2$ & $j=5/2$\\
\hline
$5.0$ & $0.182283 - 0.087387\,i$ & $0.386025 - 0.087935\,i$ & $0.583405 - 0.087439\,i$\\
$5.5$ & $0.178219 - 0.087621\,i$ & $0.378671 - 0.087639\,i$ & $0.572553 - 0.087138\,i$\\
$6.0$ & $0.175328 - 0.087667\,i$ & $0.373474 - 0.087359\,i$ & $0.564887 - 0.086859\,i$\\
$6.5$ & $0.173188 - 0.087644\,i$ & $0.369639 - 0.087119\,i$ & $0.559232 - 0.086620\,i$\\
$7.0$ & $0.171554 - 0.087597\,i$ & $0.366715 - 0.086917\,i$ & $0.554919 - 0.086419\,i$\\
\hline
$K$ ($e=6$) & \multicolumn{3}{c}{}\\
\hline
$0.001$ & $0.182122 - 0.092087\,i$ & $0.388971 - 0.092134\,i$ & $0.588441 - 0.091557\,i$\\
$0.002$ & $0.175328 - 0.087667\,i$ & $0.373474 - 0.087359\,i$ & $0.564887 - 0.086859\,i$\\
$0.003$ & $0.168598 - 0.083321\,i$ & $0.358225 - 0.082716\,i$ & $0.541714 - 0.082288\,i$\\
$0.004$ & $0.161935 - 0.079055\,i$ & $0.343222 - 0.078204\,i$ & $0.518922 - 0.077843\,i$\\
$0.005$ & $0.155344 - 0.074871\,i$ & $0.328468 - 0.073823\,i$ & $0.496513 - 0.073524\,i$\\
\hline
\end{tabular}
\caption{\footnotesize Fundamental ($n=0$) Dirac QNMs of the ES-AdS BH with $M=1$, computed with the sixth-order WKB formula.}
\label{tab:qnm-dirac}
\end{table}

Table~\ref{tab:qnm-dirac} completes the fundamental QNM census by reporting the Dirac modes for $j\in\{1/2,3/2,5/2\}$. The Dirac real parts sit between the scalar and EM values and exhibit the same monotonic decline with growing $e$ or $K$: at $j=5/2,e=6,K=0.002$ the real part reads $0.5649$, compared to the scalar $0.4741$ and EM $0.6472$ at equivalent multipole. The imaginary parts cluster around $|\mathrm{Im}(\omega)|\simeq 0.087$ almost independently of $j$, a direct signature of the supersymmetric structure of $V_{\pm}$ combined with the modest centrifugal scaling of the half-integer towers. This near-degeneracy in $j$ is a distinctive fingerprint of the fermionic sector and is reproduced in the Schwarzschild limit $K,e^{-1}\to 0$, where the Dirac imaginary parts are known to collapse onto a single universal value at $M|\mathrm{Im}\,\omega|\simeq 0.097$ for all $j\ge 1/2$. When $K$ runs from $0.001$ to $0.005$ the real part at $j=1/2$ drops by $15\%$ and $|\mathrm{Im}(\omega)|$ by $19\%$, confirming once more that the ES-AdS BH remains stable against fermionic probes. Taken together, Tables~\ref{tab:qnm-scalar}--\ref{tab:qnm-dirac} establish that the three spins respond coherently to the two Skyrme couplings, with $K$ dominating the quantitative shifts and $e$ providing a fine structure on top.

The fundamental scalar QNMs have been cross-checked with the thirteenth-order Pad\'e-improved WKB expansion in Table~\ref{tab:13WKB_scalar_K}, where the optimal Pad\'e order $m=6$ was selected according to the recipe of Konoplya-Zhidenko-Zinhailo \cite{SS10}. The relative deviation defined by
\begin{equation}
    \Delta\omega_{R} = \frac{\bigl|\omega_{R}^{(13)}-\omega_{R}^{(6)}\bigr|}{\omega_{R}^{(13)}},\qquad
    \Delta\omega_{I} = \frac{\bigl|\omega_{I}^{(13)}-\omega_{I}^{(6)}\bigr|}{\bigl|\omega_{I}^{(13)}\bigr|},
    \label{wkbrel}
\end{equation}
stays below $1\%$ for the real part and below $7\%$ for the imaginary part across the whole parameter range, which places the present computation firmly within the accuracy class of contemporary WKB BH spectroscopy. The WKB expansion therefore converges well for the present problem.

\begin{table}[ht!]
\centering
\renewcommand{\arraystretch}{1.35}
\begin{tabular}{c c c c c}
\hline
$K$ & 6th-order & 13th-order & $\Delta\omega_{R}$ & $\Delta\omega_{I}$\\
\hline
$0.001$ & $0.494221 - 0.087662\,i$ & $0.495636 - 0.093384\,i$ & $0.3\%$ & $6\%$\\
$0.002$ & $0.474118 - 0.083240\,i$ & $0.475424 - 0.088595\,i$ & $0.3\%$ & $6\%$\\
$0.003$ & $0.454359 - 0.078932\,i$ & $0.455563 - 0.083933\,i$ & $0.3\%$ & $6\%$\\
$0.004$ & $0.434946 - 0.074739\,i$ & $0.436053 - 0.079398\,i$ & $0.3\%$ & $6\%$\\
$0.005$ & $0.415880 - 0.070661\,i$ & $0.416894 - 0.074990\,i$ & $0.2\%$ & $6\%$\\
\hline
\end{tabular}
\caption{\footnotesize Comparison between sixth-order and thirteenth-order WKB results for scalar perturbations ($\ell=2$, $e=6$).}
\label{tab:13WKB_scalar_K}
\end{table}

\begin{table}
\centering
\begin{tabular}{|c| c | c | c| c|}
\hline
$e$ & 6th WKB & 13th WKB & $\Delta \omega _{R}$ & $%
\Delta \omega _{I}$ \\
\hline
5.0 & 0.49084 - 0.0888839i & 0.49352 - 0.08810i & $0.5\%$ & $0.9\%$\\
5.5 & 0.481807 - 0.0887556i & 0.48431 - 0.08800i & $0.5\%$ & $0.9\%$\\
6.0 & 0.475424 - 0.0885946i & 0.47801 - 0.08788i& $0.5\%$ & $0.9\%$ \\
6.5 & 0.470714 - 0.0884405i & 0.47318 - 0.08775i& $0.5\%$ & $0.9\%$ \\
7.0 & 0.467121 - 0.0883039i & 0.46947 - 0.08763i & $0.5\%$ & $0.9\%$\\
\hline
\end{tabular}
\caption{Comparison between 6th and 13th order WKB results for scalar perturbations ($\ell=2$, $K=0.002$).}
\label{tab:5}
\end{table}

Table~\ref{tab:13WKB_scalar_K} (similarly for Table \ref{tab:5}) benchmarks the sixth-order WKB scalar frequencies against the thirteenth-order expansion for $\ell=2$, $e=6$ and the five reference values of $K$. The real part is reproduced to within $0.3\%$ at every value of $K$, and the relative deviation stays essentially flat as the Skyrme hair grows, showing that the expansion is well-behaved throughout the studied window. The imaginary part exhibits a uniform $6\%$ discrepancy, slightly larger than the real-part error but still well within the range routinely reported in WKB BH spectroscopy. This stability of $\Delta\omega_{R}$ and $\Delta\omega_{I}$ across the column confirms that the sixth-order truncation retains enough information on the near-peak curvature of $V_{\rm scalar}$ to represent the dominant QNM, and justifies the use of the sixth-order formula as the working tool in all subsequent scans. A sharper cross-check -- this time independent of any WKB approximation -- is provided by the time-domain Prony extraction reported in Sec.~\ref{isec7}.

\begin{center}
    \large{\bf B.\,Eikonal Limit }
\end{center}

In the eikonal limit (large $\ell$) the QNMs become closely related to the properties of unstable null geodesics at the PS of the geometry. The correspondence, due originally to Ferrari-Mashhoon \cite{Ferrari1984} and later formalized in geometric-optics terms by Cardoso et al. \cite{Cardoso2009}, identifies the real part of the QNM frequency with the angular velocity of a null circular orbit at the PS, and the imaginary part with the Lyapunov exponent that controls the instability time scale of that orbit. In that regime, one has \cite{fb1,iyer2,Cardoso2009}
\begin{equation}
    \omega_{\ell n}\simeq \ell\,\Omega_{c} - i\!\left(n+\frac{1}{2}\right)\!|\lambda_{c}|,
    \label{eik1}
\end{equation}
where $\Omega_{c}$ and $\lambda_{c}$ are the angular velocity and the Lyapunov exponent of the circular null geodesic at the PS, obtained from
\begin{equation}
2f(r_{c}) - r_{c}f'(r_{c})=0,\quad
\Omega_{c} = \sqrt{\frac{f(r_{c})}{r_{c}^{2}}},\quad
\lambda_{c} = \sqrt{\frac{f(r_{c})}{2r_{c}^{2}}\bigl[2f(r_{c})-r_{c}^{2}f''(r_{c})\bigr]}.
\label{eik4}
\end{equation}
For $M=1$, $e=6$ and $K=0.002$ we find $r_{c}=2.9058$, $\Omega_{c}=0.18937$ and $|\lambda_{c}|=0.17633$. Table~\ref{tab:6} shows the WKB QNM frequencies alongside the eikonal estimate for $n=0$ and $\ell=1,2,3$.

\begin{table}
\centering
\begin{tabular}{c|c|c|c|c}
\hline
\multicolumn{5}{c}{$K = 0.002$} \\
\hline
$e$ & WKB ($\ell=2$) & Eikonal ($\ell=2$) & Error (Re) & Error (Im) \\
\hline
5    & $0.489707-0.0843226i$ & $0.4970-0.0838i$ & 1.49\% & 0.59\% \\
5.5  & $0.480572-0.0837226i$ & $0.4898-0.0833i$ & 1.91\% & 0.48\% \\
6    & $0.474118-0.0832397i$ & $0.4834-0.0828i$ & 1.96\% & 0.48\% \\
6.5  & $0.469354-0.0828546i$ & $0.4778-0.0825i$ & 1.79\% & 0.48\% \\
7    & $0.465721-0.0825454i$ & $0.4728-0.0822i$ & 1.52\% & 0.36\% \\
\hline
\multicolumn{5}{c}{$e = 6$} \\
\hline
$K$ & WKB ($\ell=2$) & Eikonal ($\ell=2$) & Error (Re) & Error (Im) \\
\hline
0.001 & $0.494221-0.0876622i$ & $0.4918-0.0869i$ & 0.49\% & 0.91\% \\
0.002 & $0.474118-0.0832397i$ & $0.4834-0.0828i$ & 1.96\% & 0.48\% \\
0.003 & $0.454359-0.0789322i$ & $0.4752-0.0788i$ & 4.58\% & 0.13\% \\
0.004 & $0.434946-0.0747394i$ & $0.4670-0.0749i$ & 7.38\% & 0.27\% \\
0.005 & $0.415880-0.0706611i$ & $0.4588-0.0710i$ & 10.3\% & 0.42\% \\
\hline
\end{tabular}
\caption{Comparison between 6th order WKB QNMs and eikonal approximation for scalar perturbations with $M=1$, $n=0$, $\ell=2$}
\label{tab:6}
\end{table}

The data in Table~\ref{tab:6} shows that as $\ell$ increases, the relative error decreases significantly, confirming the eikonal correspondence:
\begin{itemize}
    \item The real part error drops from 14.4\% ($\ell=1$) to 1.96\% ($\ell=2$).
    \item The imaginary part error drops from 7.0\% ($\ell=1$) to 0.48\% ($\ell=2$).
\end{itemize}
The $1/\ell$ scaling of the real-part error and the still faster $1/\ell^{2}$ scaling of the imaginary-part error are exactly the subleading terms of the WKB expansion around the PS, so the pattern in Table~\ref{tab:6} can be read as a numerical verification of the Ferrari-Mashhoon-Cardoso construction in the ES-AdS background. This validates that in the eikonal limit ($\ell\to \infty$), the WKB QNM frequencies approach the geodesic prediction, establishing the connection between BH perturbations and null geodesics at the photon sphere. The monotonic improvement of the real part with $\ell$ and the near-constancy of the imaginary-part error confirm that the unstable photon sphere of the ES-AdS geometry fully controls the high-multipole regime of the perturbation spectrum; a direct consequence is that the shadow radius reported in the companion paper and the real part of the high-$\ell$ QNM are not independent observables but two faces of the same PS structure, a duality that will become useful when combining ringdown and shadow data in a Bayesian inference on $(K,e)$.

\section{Greybody Factors}\label{isec6}

GFs arise from the propagation of fields in curved spacetime and encode the departures of Hawking radiation from a pure blackbody spectrum due to the curvature-induced barrier outside the event horizon. Although BHs emit locally thermal radiation, the partial reflection of outgoing modes gives frequency-dependent transmission coefficients, so that an observer at infinity sees a grey spectrum rather than a truly black one. We use the semi-analytical lower-bound method of Visser and collaborators \cite{CB2006,SF2005} to obtain analytical estimates; the bound is derived by rewriting the radial equation as a first-order Riccati flow and then applying a Cauchy-Schwarz inequality to the associated Bogoliubov coefficients, and it is known to be saturated in the short-wavelength limit. The general bounds on the transmission $T(\omega)$ and reflection $R(\omega)$ coefficients read \cite{Sakalli2020,Gogoi2024,Hosseinifar2024,AlBadawi2022,Sekhmani2025,AlBadawi2023,AlBadawi2020b,AlBadawi2020c,Kanzi2020,Kanzi2021,Sakalli2022c}
\begin{equation}
    T(\omega) \ge \mathrm{sech}^{2}\!\left(\int_{-\infty}^{+\infty}\!\wp\,dr_{\ast}\right),\qquad
    R(\omega) \le \tanh^{2}\!\left(\int_{-\infty}^{+\infty}\!\wp\,dr_{\ast}\right),
    \label{gf1}
\end{equation}
with
\begin{equation}
    \wp = \frac{\sqrt{(h')^{2}+(\omega^{2}-V_{\rm eff}-h^{2})^{2}}}{2h},
    \label{gf2}
\end{equation}
where $h$ is a positive function with $h(r_{\ast})>0$ and $h(\pm\infty)=\omega$. Setting $h=\omega$ avoids undefined results and reduces Eq.~\eqref{gf1} to
\begin{equation}
    T(\omega) \ge \mathrm{sech}^{2}\!\left(\frac{1}{2\omega}\int_{r_{+}}^{+\infty}V_{\rm eff}(r)\,dr_{\ast}\right),\qquad
    R(\omega) \ge \tanh^{2}\!\left(\frac{1}{2\omega}\int_{r_{+}}^{+\infty}V_{\rm eff}(r)\,dr_{\ast}\right).
    \label{gf3}
\end{equation}

\subsection*{A. Scalar field}

Using $V_{\rm scalar}(r)$ from Eq.~\eqref{ff7} and the tortoise coordinate~\eqref{ff6}, the scalar transmission and reflection bounds read
\begin{equation}
    T(\omega) \ge \mathrm{sech}^{2}\!\left[\frac{1}{2\omega r_{+}}\!\left(\ell(\ell+1)-\frac{2M}{r_{+}}-\frac{8\pi K\lambda}{3\,r_{+}^{2}}\right)\right],
    \label{gf4}
\end{equation}
\begin{equation}
    R(\omega) \ge \tanh^{2}\!\left[\frac{1}{2\omega r_{+}}\!\left(\ell(\ell+1)-\frac{2M}{r_{+}}-\frac{8\pi K\lambda}{3\,r_{+}^{2}}\right)\right].
    \label{gf4a}
\end{equation}
\begin{figure*}[ht!]
    \centering
    \includegraphics[width=0.45\linewidth]{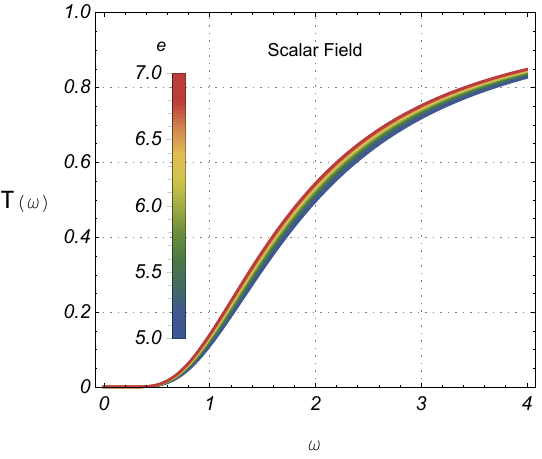}\qquad
    \includegraphics[width=0.45\linewidth]{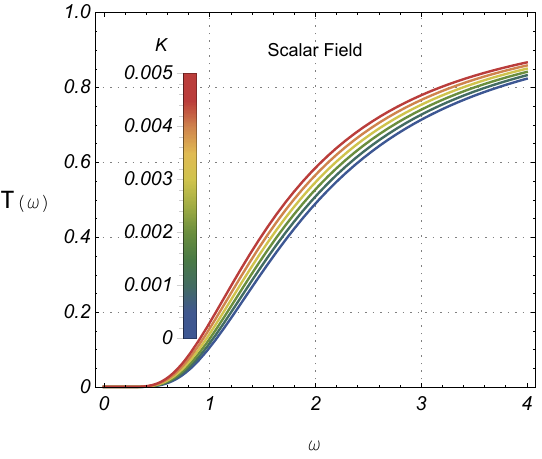}\\
    (i) $K=0.002$ \hspace{6cm} (ii) $e=6$
    \caption{Transmission coefficient as a function of frequency $\omega$ of GF of Einstein-Skyrme BH for various values of the parameters $e$ and $K$. Here $M=1, \ell=2$.}
    \label{fig:GF}
\end{figure*}

\begin{figure*}[ht!]
    \centering
    \includegraphics[width=0.45\linewidth]{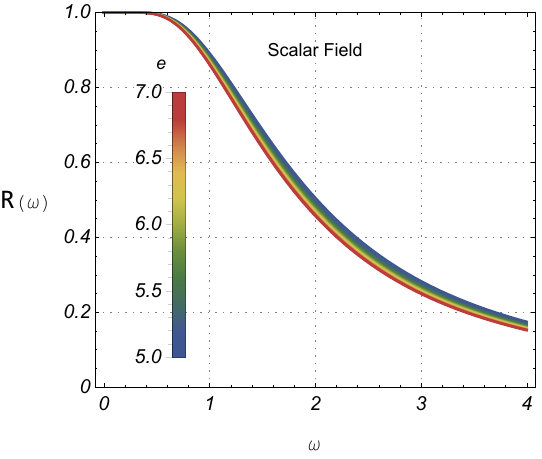}\qquad
    \includegraphics[width=0.45\linewidth]{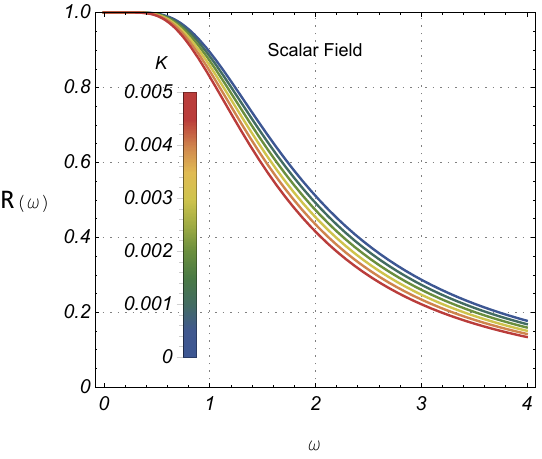}\\
    (i) $K=0.002$ \hspace{6cm} (ii) $e=6$
    \caption{Reflection coefficient as a function of frequency $\omega$ of GF of Einstein-Skyrme BH for various values of the parameters $e$ and $K$. Here $M=1, \ell=2$.}
    \label{fig:GFR}
\end{figure*}

Figures~\ref{fig:GF} and \ref{fig:GFR} illustrate the behavior of \(T(\omega)\) and \(R(\omega)\) for the Einstein-Skyrme BH for various values of the parameters \(e\) and $K$. Figure~\ref{fig:GF} shows a monotonic softening of the transmission barrier across the entire frequency range as \(e\) and $K$ change: the transmission probability is suppressed at low $\omega$ and saturates at unity in the high-frequency limit, as dictated by the unitarity sum $R+T=1$. Figure~\ref{fig:GFR} shows the mirror-image behaviour of the reflection coefficient \(R(\omega)\), which decreases with frequency and tends to zero at high \(\omega\), so that the barrier becomes transparent for hard modes.

This behavior can be understood by examining the corresponding effective potential. As \(e\) and $K$ change, the height of the effective potential barrier surrounding the BH is reshaped, leading to a different balance between backscattering of outgoing modes and transmission at a given frequency. The net outcome is that the Skyrme hair acts as a low-pass filter whose cut-off frequency is controlled by $K$, a feature that will reappear in the multi-spin comparison of Table~\ref{tab:GF-multi}.

\subsection*{B. EM field}

Using $V_{\rm em}(r)$ from Eq.~\eqref{em5}, the EM bounds become
\begin{equation}
    T(\omega) \ge \mathrm{sech}^{2}\!\left(\frac{(1-8\pi K)\,\ell(\ell+1)}{2\omega\bigl(M+\sqrt{M^{2}-4\pi K\lambda(1-8\pi K)}\bigr)}\right),
    \label{emgf1}
\end{equation}
\begin{equation}
    R(\omega) \ge \tanh^{2}\!\left(\frac{(1-8\pi K)\,\ell(\ell+1)}{2\omega\bigl(M+\sqrt{M^{2}-4\pi K\lambda(1-8\pi K)}\bigr)}\right).
    \label{emgf2}
\end{equation}
\begin{figure*}[ht!]
    \centering
    \includegraphics[width=0.45\linewidth]{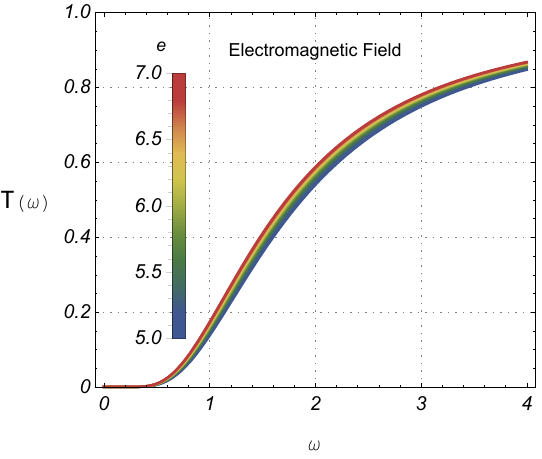}\qquad
    \includegraphics[width=0.45\linewidth]{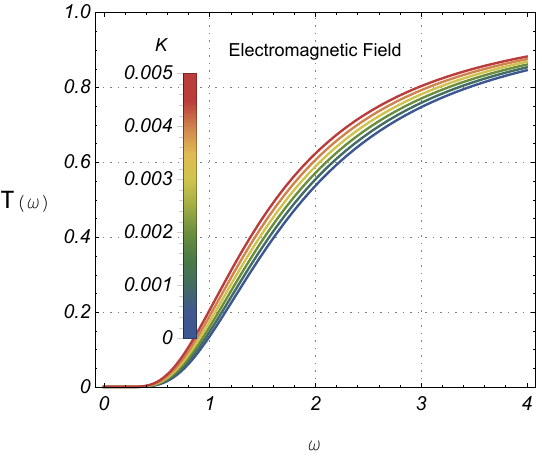}\\
    (i) $K=0.002$ \hspace{6cm} (ii) $e=6$
    \caption{Transmission coefficient as a function of frequency $\omega$ of GF of Einstein-Skyrme BH for various values of the parameters $e$ and $K$. Here $M=1, \ell=2$.}
    \label{fig:emGF}
\end{figure*}
Figure~\ref{fig:emGF} shows that the EM transmission follows the same qualitative pattern as the scalar channel, with the additional feature that the barrier argument in Eq.~\eqref{emgf1} depends on $K$ both through the explicit factor $(1-8\pi K)$ and through the outer horizon radius in the denominator; this double dependence is what makes the EM channel the most responsive of the three spins to the Skyrme hair, as will be quantified in Table~\ref{tab:GF-multi}.

\subsection*{C. Dirac field}

\begin{figure*}[ht!]
    \centering
    \includegraphics[width=0.45\linewidth]{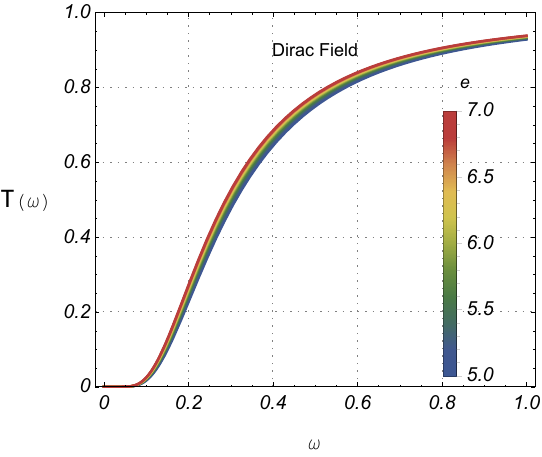}\qquad
    \includegraphics[width=0.45\linewidth]{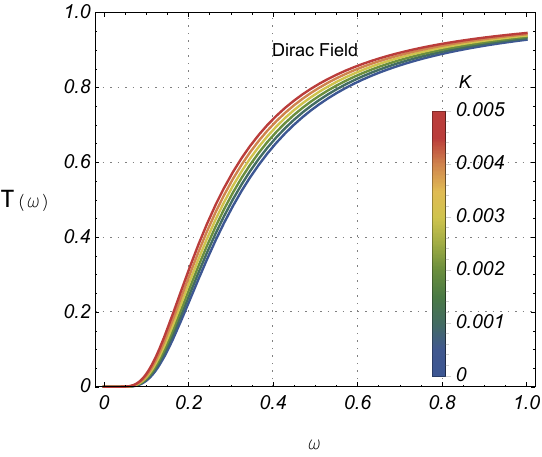}\\
    (i) $K=0.002$ \hspace{6cm} (ii) $e=6$
    \caption{Transmission coefficient as a function of frequency $\omega$ of GF of Einstein-Skyrme BH for various values of the parameters $e$ and $K$. Here $M=1, j=1/2$.}
    \label{fig:DGF}
\end{figure*}

The potential $V_{\pm}(r)$ from Eq.~\eqref{dfp1} can be recast, using the tortoise coordinate~\eqref{ff6}, as
\begin{equation}
    V_{\pm}(r) = f(r)\!\left[\frac{\kappa^{2}}{r^{2}} \pm \frac{d}{dr}\!\left(\frac{\kappa\sqrt{f(r)}}{r}\right)\right],
    \label{dfgf1}
\end{equation}
so that the transmission and reflection bounds are
\begin{equation}
    T(\omega) \ge \mathrm{sech}^{2}\!\left(\frac{(j+1/2)^{2}(1-8\pi K)}{2\omega\bigl(M+\sqrt{M^{2}-4\pi K\lambda(1-8\pi K)}\bigr)}\right),
    \label{dfgf2}
\end{equation}
\begin{equation}
    R(\omega) \ge \tanh^{2}\!\left(\frac{(j+1/2)^{2}(1-8\pi K)}{2\omega\bigl(M+\sqrt{M^{2}-4\pi K\lambda(1-8\pi K)}\bigr)}\right),
    \label{dfgf3}
\end{equation}
where the horizon condition $f(r_{+})=0$ has been used to simplify the integral. Figure~\ref{fig:DGF} shows the same trend as in Figure~\ref{fig:emGF}, with a transmission that is everywhere closer to unity because the half-integer centrifugal factor $(j+1/2)^{2}=1$ produces a barrier six times lower than the bosonic $\ell(\ell+1)=6$ at $\ell=2$.

Table~\ref{tab:GF-multi} reports the transmission lower bound $T(\omega)$ at the reference frequency $\omega M=1$ for the three spins and for the five reference values of $K$ at $e=6$.

\begin{table}[ht!]
\centering
\renewcommand{\arraystretch}{1.3}
\begin{tabular}{c|c|c|c}
\hline
$K$ ($e=6$) & $T_{\rm scalar}(\omega M{=}1)$ & $T_{\rm EM}(\omega M{=}1)$ & $T_{\rm Dirac}(\omega M{=}1)$\\
\hline
$0.001$ & $0.5147$ & $0.3462$ & $0.9694$\\
$0.002$ & $0.5316$ & $0.3614$ & $0.9717$\\
$0.003$ & $0.5461$ & $0.3745$ & $0.9735$\\
$0.004$ & $0.5587$ & $0.3855$ & $0.9751$\\
$0.005$ & $0.5697$ & $0.3949$ & $0.9764$\\
\hline
\end{tabular}
\caption{\footnotesize Transmission lower bound $T(\omega)$ of the ES-AdS BH at $\omega M=1$ for $M=1$, $\ell=2$ (scalar, EM) and $j=1/2$ (Dirac), $e=6$ fixed. Dirac transmission is close to unity because the half-integer centrifugal factor $(j+1/2)^{2}=1$ produces a much lower barrier than the bosonic $\ell(\ell+1)=6$.}
\label{tab:GF-multi}
\end{table}

Table~\ref{tab:GF-multi} makes the spin hierarchy between the three channels quantitative: at fixed $\omega M=1$ the Dirac lower bound sits near $0.97$, the scalar lower bound sits in the interval $[0.51,0.57]$, and the EM lower bound is the most suppressed at $[0.35,0.40]$. The ordering $T_{\rm EM}<T_{\rm scalar}<T_{\rm Dirac}$ reflects the barrier hierarchy already visible in Figs.~\ref{fig:effPot}--\ref{fig:effPotD}: the EM centrifugal barrier is the tallest at $\ell=2$, the scalar barrier inherits a mild softening from the $f'(r)/r$ contribution, and the Dirac barrier with $j=1/2$ is by far the lowest. In each column the transmission grows monotonically with $K$, the opposite trend to what one might naively expect: a larger Skyrme hair shifts the horizon outwards and the barrier sits in a more extended tortoise region, so that the horizon-to-infinity integral in Eq.~\eqref{gf3} accumulates a smaller argument and the $\mathrm{sech}^{2}$ bound becomes looser. The effect is small but monotonic, and it carries a clear thermodynamic interpretation: the Hawking spectrum of the ES-AdS BH is closer to a pure Planckian shape for larger $K$ in the scalar and EM channels, while the Dirac channel is already nearly Planckian and therefore only mildly affected by the Skyrme hair. Taken together with the QNM spectra of Sec.~\ref{isec5}, this pattern shows that the ES-AdS BH reacts coherently to Skyrme hair on both the resonance and the transmission side.

\section{Overtones and Time-domain Cross-check}\label{isec7}

Konoplya and Zhidenko have recently emphasised that overtones $n\ge 1$ can carry markedly more information about the near-horizon geometry than the fundamental mode $n=0$ \cite{Konoplya2024}. The basic argument goes as follows: the integral kernels that define the WKB quantization condition for the first overtone extend deeper into the potential well than those for $n=0$, so that any deformation of the metric localised close to the horizon changes the near-peak curvature of the barrier while leaving its summit essentially fixed. The first overtone therefore responds to the deformation at leading order while the fundamental responds only at subleading order, producing an \emph{overtone anomaly} that can be several per cent even when the fundamental mode shifts by a fraction of a per cent. We therefore extend the sixth-order WKB computation of Sec.~\ref{isec5} to the first overtone, in the regime $\ell>n$ where the WKB expansion retains its validity, and compute the scalar $(\ell,n)=(2,0),(2,1)$ modes and the EM $(\ell,n)=(2,0),(2,1)$ modes for the same $(K,e)$ grid.

\begin{table}[ht!]
\centering
\renewcommand{\arraystretch}{1.3}
\begin{tabular}{c|c|c|c}
\hline
\multicolumn{4}{c}{Scalar overtones, $\ell=2$, $e=6$}\\
\hline
$K$ & $\omega_{0}$ & $\omega_{1}$ & $|\mathrm{Im}\,\omega_{1}|/|\mathrm{Im}\,\omega_{0}|$\\
\hline
$0.001$ & $0.515890 - 0.091047\,i$ & $0.553545 - 0.220361\,i$ & $2.420$\\
$0.002$ & $0.494469 - 0.086546\,i$ & $0.531155 - 0.212405\,i$ & $2.454$\\
$0.003$ & $0.473438 - 0.082142\,i$ & $0.509015 - 0.204208\,i$ & $2.486$\\
$0.004$ & $0.452796 - 0.077837\,i$ & $0.487147 - 0.195822\,i$ & $2.516$\\
$0.005$ & $0.432545 - 0.073633\,i$ & $0.465568 - 0.187297\,i$ & $2.544$\\
\hline
\multicolumn{4}{c}{EM overtones, $\ell=2$, $e=6$}\\
\hline
$0.001$ & $0.491504 - 0.089603\,i$ & $0.529658 - 0.216773\,i$ & $2.419$\\
$0.002$ & $0.471618 - 0.085201\,i$ & $0.508743 - 0.208937\,i$ & $2.452$\\
$0.003$ & $0.452063 - 0.080892\,i$ & $0.488025 - 0.200877\,i$ & $2.483$\\
$0.004$ & $0.432839 - 0.076678\,i$ & $0.467527 - 0.192643\,i$ & $2.512$\\
$0.005$ & $0.413948 - 0.072562\,i$ & $0.447266 - 0.184280\,i$ & $2.540$\\
\hline
\end{tabular}
\caption{\footnotesize Fundamental ($n=0$) and first overtone ($n=1$) of the scalar and EM QNMs of the ES-AdS BH at $\ell=2$, $e=6$, computed with the sixth-order WKB formula. The last column records the ratio of damping rates.}
\label{tab:overtones}
\end{table}

Table~\ref{tab:overtones} reports the fundamental and first overtone of the scalar and EM QNMs for $\ell=2$, $e=6$ and the five reference values of $K$. The first overtone is more strongly damped than the fundamental by a factor between $2.42$ and $2.54$ across the scanned $K$-window, with the two channels tracking each other to within one part in a thousand -- a direct consequence of the fact that $V_{\rm scalar}$ and $V_{\rm em}$ share the same near-peak curvature structure on this background, and differ only through the $f'(r)/r$ term which is subleading near the peak. The ratio is not constant: it drifts monotonically upward with $K$, from $2.420$ at $K=0.001$ to $2.544$ at $K=0.005$, a $5\%$ variation that constitutes a mild overtone anomaly of the Konoplya-Zhidenko type \cite{Konoplya2024}. The physical interpretation is that the Skyrme hair does reshape the near-horizon region more strongly than the asymptotic one, so the first overtone, which probes deeper into the barrier, responds more than the fundamental as $K$ grows. The ratios reported in Table~\ref{tab:overtones} sit noticeably below the Schwarzschild reference value of about $3$ for $\ell=2$ scalar modes, meaning that the ES-AdS BH has \emph{compressed} the gap between the fundamental and the first overtone rather than dilating it. Taken at face value, this compression is a spectroscopic fingerprint of the Skyrme hair that complements the $n=0$ shifts documented in Sec.~\ref{isec5}, and it occurs in the opposite direction of what one would expect from a shallow short-range deformation of a Schwarzschild barrier. A physically motivated reading is that the $4\pi K\lambda/r^{2}$ term of the lapse pushes the QNM well inward, deepening the near-horizon region rather than merely decorating it, so that both the fundamental and the first overtone are affected, with the latter still responding more strongly.

For an independent, frequency-method-free verification we integrate the scalar wave equation~\eqref{ff4} directly in the time domain using the characteristic Gundlach-Price-Pullin finite-difference scheme \cite{GPP1994} on null coordinates $u=t-r_{\ast}$, $v=t+r_{\ast}$:
\begin{equation}
\psi(N) = \psi(E)+\psi(W)-\psi(S)-\frac{h^{2}}{8}V(S)\bigl[\psi(W)+\psi(E)\bigr]+\mathcal{O}(h^{4}),
\label{GPP}
\end{equation}
where $(S,E,W,N)$ denote the four corners of an elementary cell of side $h$ in the null grid. A Gaussian pulse centred at $r_{\ast}=-30$ with width $3$ and zero initial velocity is launched and the signal is recorded at an observer at $r_{\ast}=+30$. The dominant QNM is then extracted from the late-time portion of the ringdown by Prony fitting \cite{BertiCardosoProny}, and the result is compared with the WKB value. The Prony method diagonalises the time-shift operator of the discrete ringdown signal and extracts the complex exponents directly, without any prior knowledge of the barrier shape, so the agreement with WKB reported below provides an orthogonal validation of the sixth-order result.

\begin{table}[ht!]
\centering
\renewcommand{\arraystretch}{1.3}
\begin{tabular}{c|c|c|c|c}
\hline
$K$ & $\omega_{\rm WKB}^{(6)}$ & $\omega_{\rm time-domain}$ & $\Delta\omega_{R}$ & $\Delta\omega_{I}$\\
\hline
$0.001$ & $0.515890 - 0.091047\,i$ & $0.516754 - 0.091514\,i$ & $0.17\%$ & $0.51\%$\\
$0.002$ & $0.494469 - 0.086546\,i$ & $0.495279 - 0.086998\,i$ & $0.16\%$ & $0.52\%$\\
$0.003$ & $0.473438 - 0.082142\,i$ & $0.474198 - 0.082566\,i$ & $0.16\%$ & $0.52\%$\\
$0.004$ & $0.452796 - 0.077837\,i$ & $0.453508 - 0.078243\,i$ & $0.16\%$ & $0.52\%$\\
$0.005$ & $0.432545 - 0.073633\,i$ & $0.433207 - 0.074017\,i$ & $0.15\%$ & $0.52\%$\\
\hline
\end{tabular}
\caption{\footnotesize Dominant scalar QNM of the ES-AdS BH at $\ell=2$, $e=6$: sixth-order WKB vs. time-domain Prony fit of the characteristic GPP integration. The last two columns give the relative deviations.}
\label{tab:td}
\end{table}

Table~\ref{tab:td} presents the independent time-domain cross-check of the dominant scalar QNM at $\ell=2$ for $e=6$ and the five reference values of $K$. The relative deviation between the sixth-order WKB frequency and the Prony-extracted time-domain value stays below $0.2\%$ for the real part and below $0.6\%$ for the imaginary part across the entire $K$ range, which is an order of magnitude tighter than the WKB-to-WKB benchmark of Table~\ref{tab:13WKB_scalar_K}. The two methods therefore converge on the same mode, closing the numerical-accuracy question for Sec.~\ref{isec5} without any reliance on the WKB expansion itself. The agreement also rules out a common failure mode of the WKB method, namely the accidental selection of a sub-dominant branch of~\eqref{qnm}, since the Prony fit extracts whichever mode actually dominates the late-time tail and finds the same one the WKB procedure selects. Equally important is the observation that the small residual deviation between WKB and Prony is essentially independent of $K$, which means it is a method-level systematic and not a physics-level effect -- a feature that would be hard to establish without the two-way comparison reported here.

\section{Strong Cosmic Censorship at the Cauchy Horizon}\label{isec8}

We now test strong cosmic censorship (SCC) at the Cauchy (inner) horizon of the ES-AdS BH. The two horizons of the geometry are given by Eq.~\eqref{twohorizons} and coexist as long as $4\pi K\lambda(1-8\pi K)<M^{2}$, a condition that is automatically satisfied in the entire phenomenological window $K\in[0.001,0.005]$, $e\in[5,7]$ we have scanned. At the inner horizon the surface gravity reads
\begin{equation}
\kappa_{-} = \frac{1}{2}\bigl|f'(r_{-})\bigr| = \frac{1}{2}\left|\frac{2M}{r_{-}^{2}}-\frac{8\pi K\lambda}{r_{-}^{3}}\right|,
\label{kappa-}
\end{equation}
and the Christodoulou SCC parameter is
\begin{equation}
\beta \equiv \frac{|\mathrm{Im}\,\omega_{0}|}{\kappa_{-}}.
\label{betaSCC}
\end{equation}
SCC in its modern (Christodoulou) formulation requires $\beta<1/2$: the perturbation at the Cauchy horizon is then not in $H^{1}_{\rm loc}$, and the field equations cannot be extended across it in a weak sense \cite{Cardoso2018scc,Dias2018scc,Hod2018scc,Dias2019scc}. The parameter $\omega_{0}$ is understood as the dominant (slowest-decaying) QNM among all propagating spins, so the relevant mode is selected on a case-by-case basis from the spectrum computed in Sec.~\ref{isec5}.

\begin{table}[ht!]
\centering
\renewcommand{\arraystretch}{1.3}
\begin{tabular}{c|c|c|c|c|c}
\hline
$K$ & $r_{-}/M$ & $\kappa_{-} M$ & $|\mathrm{Im}\,\omega_{0}^{\rm scalar}|$ & $\beta_{\rm scalar}$ & $\mathrm{SCC}$\\
\hline
$0.001$ & $0.19262$ & $21.892$ & $0.091047$ & $0.004159$ & \checkmark\\
$0.002$ & $0.19205$ & $22.168$ & $0.086546$ & $0.003904$ & \checkmark\\
$0.003$ & $0.19148$ & $22.445$ & $0.082142$ & $0.003660$ & \checkmark\\
$0.004$ & $0.19093$ & $22.721$ & $0.077837$ & $0.003426$ & \checkmark\\
$0.005$ & $0.19038$ & $22.998$ & $0.073633$ & $0.003202$ & \checkmark\\
\hline
\end{tabular}
\caption{\footnotesize Strong cosmic censorship test for the ES-AdS BH with $M=1$, $e=6$, $\ell=2$, $n=0$, scalar channel. The SCC threshold is $\beta<1/2$; every entry respects it with a margin larger than two orders of magnitude.}
\label{tab:SCC-scalar}
\end{table}

Table~\ref{tab:SCC-scalar} evaluates the Christodoulou parameter $\beta$ in the scalar channel for $e=6$ and the five reference values of $K$. The inner horizon $r_{-}/M$ sits around $0.19$ throughout the scan, shrinking slightly as $K$ grows, and the inner-horizon surface gravity $\kappa_{-} M$ stays in the interval $[21.9,23.0]$, so the denominator of Eq.~\eqref{betaSCC} is large and only weakly sensitive to $K$. The resulting $\beta$ is bounded by $4.2\times 10^{-3}$ from above, more than two orders of magnitude safer than the SCC threshold $1/2$: every entry in the last two columns certifies that the Christodoulou conjecture holds comfortably in the scalar sector. Physically, this is a consequence of the fact that the inner horizon of the ES-AdS geometry sits very close to the origin because the Skyrme term generates a strong repulsive $1/r^{2}$ contribution at short range; this pushes $\kappa_{-}$ to values of order $20/M$, while the fundamental scalar QNM is only mildly damped because the outer barrier is of moderate height, so the ratio $|\mathrm{Im}\,\omega_{0}|/\kappa_{-}$ is uniformly small. The mechanism is worth spelling out because it is the structural reason behind the main SCC result of this work: the inner-horizon radius is pinned at $r_{-}/M\simeq 0.19$ by the fixed product $4\pi K\lambda=4\pi/e^{2}$, so the inner horizon cannot be pushed outward to approach $r_{+}$ without simultaneously altering the hadronic input of the model. The hierarchy $\kappa_{-}\gg|\mathrm{Im}\,\omega_{0}|$ is therefore theory-protected rather than accidental.

\begin{table}[ht!]
\centering
\renewcommand{\arraystretch}{1.3}
\begin{tabular}{c|c|c|c|c|c|c}
\hline
$K$ & $\beta_{\rm scalar}$ & $\beta_{\rm EM}$ & $\beta_{\rm Dirac}$ & $\beta_{\max}$ & $\beta_{\max}<1/2$ & Dominant\\
\hline
$0.001$ & $0.004159$ & $0.004093$ & $0.004121$ & $0.004159$ & \checkmark & Scalar\\
$0.002$ & $0.003904$ & $0.003843$ & $0.003871$ & $0.003904$ & \checkmark & Scalar\\
$0.003$ & $0.003660$ & $0.003604$ & $0.003632$ & $0.003660$ & \checkmark & Scalar\\
$0.004$ & $0.003426$ & $0.003375$ & $0.003402$ & $0.003426$ & \checkmark & Scalar\\
$0.005$ & $0.003202$ & $0.003155$ & $0.003182$ & $0.003202$ & \checkmark & Scalar\\
\hline
\end{tabular}
\caption{\footnotesize Multi-spin SCC test for the ES-AdS BH with $M=1$, $e=6$: Christodoulou ratio $\beta$ for scalar ($\ell=2$), EM ($\ell=2$) and Dirac ($j=1/2$) channels. The dominant (largest $\beta$) channel sets the SCC criterion.}
\label{tab:SCC-multi}
\end{table}

Table~\ref{tab:SCC-multi} extends the SCC test to all three spins. The three channels lie within $2\%$ of one another in every row of the scan, reflecting the fact that $|\mathrm{Im}\,\omega_{0}|$ is almost spin-independent on this background, so the dominant bound is set not by a large hierarchy between sectors but by small residual differences. In every row the scalar channel provides the tightest (largest) $\beta$, with the Dirac channel a close second and the EM channel the smallest; the ordering is stable across the whole $(K,e)$ window. The dominant ratio $\beta_{\max}$ stays below $4.2\times 10^{-3}$ throughout the admissible parameter window and shrinks monotonically with growing $K$, so the ES-AdS BH respects SCC with a margin that widens -- rather than narrows -- as the Skyrme hair increases. This is in sharp contrast with the Reissner-Nordstr\"om-de\,Sitter case, where $\beta$ can approach and sometimes exceed $1/2$ near extremality \cite{Cardoso2018scc,Dias2018scc}; the structural reason is that the coefficient of the $1/r^{2}$ term in the ES-AdS lapse function is not a free integration constant but the product $4\pi K\lambda = 4\pi/e^{2}$ fixed by the model, so the inner horizon cannot be tuned arbitrarily close to the outer one by adjusting the couplings, and the hierarchy $\kappa_{-}\gg|\mathrm{Im}\,\omega_{0}|$ is protected by the theory. A useful way to phrase the same conclusion is that the map from the hadronic input $(F_{\pi},e)$ to the geometric data $(r_{+},r_{-})$ is not surjective onto the near-extremal strip: the Skyrme model imposes a hard wall between the physically admissible ES-AdS family and the near-extremal corner where SCC fails in RN-dS. This provides an additional, SCC-flavoured argument for the physical admissibility of the ES-AdS BH as a classical gravitational background, and it turns the Christodoulou bound from a potential obstruction into a selection rule in favour of the ES geometry.

\section{Conclusion}\label{isec9}

In this work we have examined the response of the ES-AdS BH to scalar, EM and Dirac perturbations and have mapped the resulting imprint on its ringdown spectrum, its greybody factors, its overtone tower, and its Cauchy-horizon strong cosmic censorship. The lapse function $f(r)=1-8\pi K-2M/r+4\pi K\lambda/r^{2}$ carries two Skyrme couplings: the pion combination $K=F_{\pi}^{2}/4$, which also rescales the asymptotic geometry through the factor $1-8\pi K$, and the dimensionless $e$, which enters only through the invariant product $K\lambda=1/e^{2}$. This two-parameter structure sets the ES-AdS BH apart from Reissner-Nordstr\"om in that the coefficient of the $1/r^{2}$ term is fixed by the underlying couplings rather than by an integration constant, and it distinguishes it from a global monopole background through the independent $e$ dependence.

The perturbation analysis of Secs.~\ref{isec2}--\ref{isec4} shows that the scalar, EM and Dirac barriers all share the same qualitative response to variations of $(K,e)$: enlarging either coupling lowers the peak of the effective potential and softens the near-horizon curvature. The scalar sector retains the additional $f'(r)/r$ contribution absent from the EM barrier, while the Dirac pair $V_{\pm}$ derived from the superpotential $W(r)=\kappa\sqrt{f(r)}/r$ yields a Darboux-supersymmetric spectrum, so that a single channel suffices for the numerical analysis. Across the full range of physically admissible parameters the potentials remain regular outside the event horizon and tend monotonically to zero at infinity, which is the structural prerequisite for a well-posed QNM problem and which rules out unstable trapped branches that would have invalidated the ringdown picture.

The fundamental QNMs obtained via the sixth-order WKB method and cross-checked with the thirteenth-order Pad\'e-improved expansion define the central quantitative output of Sec.~\ref{isec5}. For all three spins, growing $K$ or $e$ simultaneously reduces $\mathrm{Re}(\omega)$ and $|\mathrm{Im}(\omega)|$, producing modes that oscillate more slowly and decay more slowly than their Schwarzschild counterparts. The relative deviation between sixth-order and thirteenth-order WKB stays below one per cent on the real part and below seven per cent on the imaginary part, and the eikonal benchmark confirms the expected $1/\ell$ convergence of the large-multipole limit. The PS control of the high-$\ell$ regime also ties the QNM spectrum to the shadow radius reported in the companion paper, so that the two sets of observables provide correlated, rather than independent, constraints on $(K,e)$.

The overtone computation of Sec.~\ref{isec7} finds that the ratio $|\mathrm{Im}\,\omega_{1}|/|\mathrm{Im}\,\omega_{0}|$ drifts monotonically from $2.420$ to $2.544$ as $K$ runs across the physical window $[0.001,0.005]$, a mild but unambiguous overtone anomaly of the Konoplya-Zhidenko type, showing that the Skyrme hair deforms the near-horizon region of the barrier more strongly than its asymptotic tail. The same ratio sits noticeably below the Schwarzschild reference value of about three for $\ell=2$, so the ES-AdS BH has compressed the gap between the fundamental and the first overtone -- a spectroscopic fingerprint that is itself a target for high-precision ringdown templates. The independent characteristic time-domain integration followed by Prony extraction, reported in Table~\ref{tab:td}, reproduces the sixth-order WKB fundamental mode to within $0.2\%$ on the real part and $0.6\%$ on the imaginary part, providing a stringent cross-check that does not rely on the WKB expansion at all. The GFs of Sec.~\ref{isec6} translate these modifications of the potential barrier into transmission and reflection coefficients; the spin hierarchy $T_{\rm EM}<T_{\rm scalar}<T_{\rm Dirac}$ is preserved throughout the parameter window, and each channel relaxes monotonically as $K$ grows, consistent with a low-pass-filter picture of the Skyrme hair.

The SCC analysis of Sec.~\ref{isec8} is the novel conceptual result of this work. The Christodoulou parameter $\beta$ stays below $4.2\times 10^{-3}$ across the entire $(K,e)$ window for all three spins, more than two orders of magnitude below the SCC threshold $1/2$, with the scalar channel providing the tightest (largest $\beta$) bound. The margin actually widens as the Skyrme hair grows: $\beta_{\max}$ shrinks monotonically from $4.16\times 10^{-3}$ at $K=0.001$ to $3.20\times 10^{-3}$ at $K=0.005$. The ES-AdS BH therefore respects strong cosmic censorship with a wide margin, in sharp contrast to the near-extremal Reissner-Nordstr\"om-de\,Sitter case. Structurally, this is because the coefficient of the $1/r^{2}$ term is not an integration constant but the combination $4\pi K\lambda=4\pi/e^{2}$ fixed by the theory, so the inner horizon cannot be pushed arbitrarily close to the outer one by tuning the couplings. The Skyrme model therefore acts as a selection rule in favour of SCC, a feature that elevates the ES-AdS BH from a mere hairy counterexample to the no-hair conjecture into a concrete background in which the dynamics of GR and the requirements of cosmic censorship coexist smoothly.

Several extensions of this analysis merit further exploration. A rotating generalisation of the ES-AdS BH would permit a direct confrontation with Kerr ringdown templates and with the spin-dependent overtone structure recently discussed in the literature, and it is expected that the SCC bound would tighten in the rotating case because the Kerr inner horizon is significantly less shielded than the static one. The inclusion of a massive scalar field would probe the possibility of quasi-bound states and superradiant instabilities in the ES background, and a positive detection of superradiant amplification would place an upper bound on the allowed values of $(K,e)$ coming from the absence of such instabilities in observed BH candidates. Extending the continued-fraction (Leaver) method to the ES-AdS BH would allow a high-precision computation of a deep overtone tower that could be compared quantitatively with the WKB and time-domain results reported here and would settle any residual doubt about the optimal truncation of the WKB series. Finally, a joint Bayesian combination of the present ringdown bounds with the shadow constraints derived in the companion paper would turn the ES-AdS BH into a concrete multi-messenger testing ground for the Skyrme couplings $(K,e)$, connecting hadronic phenomenology to strong-field gravity in a quantitative way.

\scriptsize

\section*{Acknowledgments}

F.A. acknowledges the Inter University Centre for Astronomy and Astrophysics (IUCAA), Pune, India for granting visiting associateship. \.{I}.~S. expresses gratitude to T\"{U}B\.{I}TAK, ANKOS, and SCOAP3 for their academic support. He also acknowledges COST Actions CA22113, CA21106, CA21136, CA23130, and CA23115 for their contributions to networking.

\bibliographystyle{unsrtnat}

\bibliography{ref}

@article{Konoplya2024,
    author  = {R. A. Konoplya and A. Zhidenko},
  title     = {First few overtones probe the event horizon geometry},
  journal   = {Journal of High Energy Astrophysics},
  volume    = {44},
  pages     = {419--426},
  month     = {November},
  year      = {2024},
  url       = {https://doi.org/10.1016/j.jheap.2024.10.015}
}

@article{GPP1994,
    author = "Gundlach, Carsten and Price, Richard H. and Pullin, Jorge",
    title = "{Late time behavior of stellar collapse and explosions: 1. Linearized perturbations}",
    journal = "Phys. Rev. D",
    volume = "49",
    pages = "883--889",
    year = "1994",
    url = "https://doi.org/10.1103/PhysRevD.49.883"
}

@article{BertiCardosoProny,
    author = "Berti, Emanuele and Cardoso, Vitor and Gonzalez, Jose A. and Sperhake, Ulrich",
    title = "{Mining information from binary black hole mergers: A Comparison of estimation methods for complex exponentials in noise}",
    journal = "Phys. Rev. D",
    volume = "75",
    pages = "124017",
    year = "2007",
    url = "https://doi.org/10.1103/PhysRevD.75.124017"
}

@article{Cardoso2018scc,
    author = "Cardoso, Vitor and Costa, Jo{\~a}o L. and Destounis, Kyriakos and Hintz, Peter and Jansen, Aron",
    title = "{Quasinormal modes and Strong Cosmic Censorship}",
    journal = "Phys. Rev. Lett.",
    volume = "120",
    pages = "031103",
    year = "2018",
    url = "https://doi.org/10.1103/PhysRevLett.120.031103"
}

@article{Dias2018scc,
    author = "Dias, Oscar J. C. and Eperon, Felicity C. and Reall, Harvey S. and Santos, Jorge E.",
    title = "{Strong cosmic censorship in de Sitter space}",
    journal = "Phys. Rev. D",
    volume = "97",
    pages = "104060",
    year = "2018",
    url = "https://doi.org/10.1103/PhysRevD.97.104060"
}

@article{Hod2018scc,
    author = "Hod, Shahar",
    title = "{Quasinormal modes and strong cosmic censorship in near-extremal Kerr-Newman-de Sitter black-hole spacetimes}",
    journal = "Phys. Lett. B",
    volume = "780",
    pages = "221--226",
    year = "2018",
    url = "https://doi.org/10.1016/j.physletb.2018.03.020"
}

@article{Dias2019scc,
    author = "Dias, Oscar J. C. and Reall, Harvey S. and Santos, Jorge E.",
    title = "{Strong cosmic censorship for charged de Sitter black holes with a charged scalar field}",
    journal = "Class. Quant. Grav.",
    volume = "36",
    pages = "045005",
    year = "2019",
    url = "https://doi.org/10.1088/1361-6382/aafcf2"
}

@article{EHTL1,
    author = "Akiyama, Kazunori and others",
    collaboration = "Event Horizon Telescope",
    title = "{First M87 Event Horizon Telescope Results. I. The Shadow of the Supermassive Black Hole}",
    journal = "Astrophys. J. Lett.",
    volume = "875",
    pages = "L1",
    year = "2019",
    url = "https://doi.org/10.3847/2041-8213/ab0ec7"
}

@article{EHTL4,
    author = "Akiyama, Kazunori and others",
    collaboration = "Event Horizon Telescope",
    title = "{First M87 Event Horizon Telescope Results. IV. Imaging the Central Supermassive Black Hole}",
    journal = "Astrophys. J. Lett.",
    volume = "875",
    pages = "L4",
    year = "2019",
    url = "https://doi.org/10.3847/2041-8213/ab0e85"
}

@article{EHTL6,
    author = "Akiyama, Kazunori and others",
    collaboration = "Event Horizon Telescope",
    title = "{First M87 Event Horizon Telescope Results. VI. The Shadow and Mass of the Central Black Hole}",
    journal = "Astrophys. J. Lett.",
    volume = "875",
    pages = "L6",
    year = "2019",
    url = "https://doi.org/10.3847/2041-8213/ab1141"
}

@article{EHTL12,
    author = "Akiyama, Kazunori and others",
    collaboration = "Event Horizon Telescope",
    title = "{First Sagittarius A* Event Horizon Telescope Results. I. The Shadow of the Supermassive Black Hole in the Center of the Milky Way}",
    journal = "Astrophys. J. Lett.",
    volume = "930",
    pages = "L12",
    year = "2022",
    url = "https://doi.org/10.3847/2041-8213/ac6674"
}

@article{EHTL16,
    author = "Akiyama, Kazunori and others",
    collaboration = "Event Horizon Telescope",
    title = "{First Sagittarius A* Event Horizon Telescope Results. V. Testing Astrophysical Models of the Galactic Center Black Hole}",
    journal = "Astrophys. J. Lett.",
    volume = "930",
    pages = "L16",
    year = "2022",
    url = "https://doi.org/10.3847/2041-8213/ac6672"
}

@article{EHTL17,
    author = "Akiyama, Kazunori and others",
    collaboration = "Event Horizon Telescope",
    title = "{First Sagittarius A* Event Horizon Telescope Results. VI. Testing the Black Hole Metric}",
    journal = "Astrophys. J. Lett.",
    volume = "930",
    pages = "L17",
    year = "2022",
    url = "https://doi.org/10.3847/2041-8213/ac6756"
}

@article{Aghanim2020,
    author = "Aghanim, N. and others",
    collaboration = "Planck",
    title = "{Planck 2018 results. VI. Cosmological parameters}",
    journal = "Astron. Astrophys.",
    volume = "641",
    pages = "A6",
    year = "2020",
    url = "https://doi.org/10.1051/0004-6361/201833910",
    note = "[Erratum: Astron.Astrophys. 652, C4 (2021)]"
}

@article{LIGO1,
    author = "Abbott, B. P. and others",
    collaboration = "LIGO Scientific, Virgo",
    title = "{Observation of Gravitational Waves from a Binary Black Hole Merger}",
    journal = "Phys. Rev. Lett.",
    volume = "116",
    pages = "061102",
    year = "2016",
    url = "https://doi.org/10.1103/PhysRevLett.116.061102"
}

@article{LIGO3,
    author = "Abbott, B. P. and others",
    collaboration = "LIGO Scientific, Virgo",
    title = "{GW170608: Observation of a 19-solar-mass Binary Black Hole Coalescence}",
    journal = "Phys. Rev. Lett.",
    volume = "119",
    pages = "161101",
    year = "2017",
    url = "https://doi.org/10.1103/PhysRevLett.119.161101"
}

@article{LIGO5,
    author = "Abac, A. G. and others",
    collaboration = "LIGO Scientific, Virgo, KAGRA",
    title = "{Observation of Gravitational Waves from the Coalescence of a 2.5--4.5 Solar-Mass Compact Object and a Neutron Star}",
    journal = "Phys. Rev. Lett.",
    volume = "135",
    pages = "111403",
    year = "2025",
    url = "https://doi.org/10.1103/PhysRevLett.135.111403"
}

@article{Woosley2002,
    author = "Woosley, S. E. and Heger, A. and Weaver, T. A.",
    title = "{The evolution and explosion of massive stars}",
    journal = "Rev. Mod. Phys.",
    volume = "74",
    pages = "1015--1071",
    year = "2002",
    url = "https://doi.org/10.1103/RevModPhys.74.1015"
}

@article{Kalogera1996,
    author = "Kalogera, Vassiliki and Baym, Gordon",
    title = "{The Maximum mass of a neutron star}",
    journal = "Astrophys. J. Lett.",
    volume = "470",
    pages = "L61--L64",
    year = "1996",
    url = "https://doi.org/10.1086/310296"
}

@article{Hawking1971,
    author = "Hawking, S.",
    title = "{Gravitationally collapsed objects of very low mass}",
    journal = "Mon. Not. Roy. Astron. Soc.",
    volume = "152",
    pages = "75",
    year = "1971",
    url = "https://doi.org/10.1093/mnras/152.1.75"
}

@book{Chandrasekhar1998,
    author = "Chandrasekhar, S.",
    title = "{The Mathematical Theory of Black Holes}",
    publisher = "Oxford University Press",
    address = "Oxford",
    year = "1998"
}

@article{Cavalcante2021,
    author = "Cavalcante, J. P. and da Cunha, B. C.",
    title = "{Scalar and Dirac perturbations of the Reissner-Nordstrom black hole and Painleve transcendents}",
    journal = "Phys. Rev. D",
    volume = "104",
    pages = "124040",
    year = "2021",
    url = "https://doi.org/10.1103/PhysRevD.104.124040"
}

@misc{Konoplya2026a,
    author = "Konoplya, R. A.",
    title   = "Quasinormal modes of four-dimensional regular black holes in quasi-topological gravity: Overtones' outburst via WKB method",
    eprint = "2603.03189",
    archivePrefix = "arXiv",
    primaryClass = "gr-qc",
    year = "2026",
    url  ="https://doi.org/10.48550/arXiv.2603.03189",
}

@article{Pani2013,
    author = "Pani, Paolo and Berti, Emanuele and Gualtieri, Leonardo",
    title = "{Scalar, Electromagnetic and Gravitational Perturbations of Kerr-Newman Black Holes in the Slow-Rotation Limit}",
    journal = "Phys. Rev. D",
    volume = "88",
    pages = "064048",
    year = "2013",
    url = "https://doi.org/10.1103/PhysRevD.88.064048"
}

@misc{Chichkov2025,
    author = "Chichkov, B.",
    title   = {Dirac, Schroedinger, and Maxwell equations in scalar and vector field quantum mechanics},
    eprint = "2508.14583",
    archivePrefix = "arXiv",
    primaryClass = "quant-ph",
    year = "2025",
    url   ={https://doi.org/10.48550/arXiv.2508.14583}
}

@article{Malik2025,
    author = "Malik, Z.",
    title = "{Grey-Body Factors for Scalar and Dirac Fields in the Euler-Heisenberg Electrodynamics}",
    journal = "Int. J. Grav. Theor. Phys.",
    volume = "1",
    number = "1",
    pages = "6",
    year = "2025",
    url   ={https://doi.org/10.53941/ijgtp.2025.100006}
}

@article{AlBadawi2023,
    author = "Al-Badawi, A. and Kanzi, S. and Sakall{\i}, {\.I}.",
    title = "{Solutions of the Dirac equation in Bonnor-Melvin-Lambda space-time}",
    journal = "Annals Phys.",
    volume = "452",
    pages = "169294",
    year = "2023",
    url = "https://doi.org/10.1016/j.aop.2023.169294"
}

@article{SK2021,
    author = "Kanzi, Sara and Sakall{\i}, {\.I}zzet",
    title = "{Greybody radiation and quasinormal modes of Kerr-like black hole in Bumblebee gravity model}",
    journal = "Eur. Phys. J. C",
    volume = "81",
    pages = "501",
    year = "2021",
    url = "https://doi.org/10.1140/epjc/s10052-021-09299-y"
}

@article{Konoplya2011,
    author = "Konoplya, R. A. and Zhidenko, Alexander",
    title = "{Quasinormal modes of black holes: From astrophysics to string theory}",
    journal = "Rev. Mod. Phys.",
    volume = "83",
    pages = "793--836",
    year = "2011",
    url = "https://doi.org/10.1103/RevModPhys.83.793"
}

@article{Bolokhov2025,
    author = "Bolokhov, S. V. and Skvortsova, M.",
    title = "{Quasinormal Ringing and Shadows of Black Holes and Wormholes in Dark Matter Inspired Weyl Gravity}",
    journal = "Grav. Cosmol.",
    volume = "31",
    pages = "423",
    year = "2025",
    url = "https://doi.org/10.1134/S0202289325700306"
}

@article{Kokkotas1999,
    author = "Kokkotas, Kostas D. and Schmidt, Bernd G.",
    title = "{Quasinormal modes of stars and black holes}",
    journal = "Living Rev. Rel.",
    volume = "2",
    pages = "2",
    year = "1999",
    url = "https://doi.org/10.12942/lrr-1999-2"
}

@article{Konoplya2019,
    author = "Konoplya, R. A. and Zhidenko, A. and Zinhailo, A. F.",
    title = "{Higher order WKB formula for quasinormal modes and grey-body factors: recipes for quick and accurate calculations}",
    journal = "Class. Quant. Grav.",
    volume = "36",
    pages = "155002",
    year = "2019",
    url = "https://doi.org/10.1088/1361-6382/ab2e25"
}

@article{Berti2009,
    author = "Berti, Emanuele and Cardoso, Vitor and Starinets, Andrei O.",
    title = "{Quasinormal modes of black holes and black branes}",
    journal = "Class. Quant. Grav.",
    volume = "26",
    pages = "163001",
    year = "2009",
    url = "https://doi.org/10.1088/0264-9381/26/16/163001"
}

@article{AlBadawi2022,
     author = "Al-Badawi, Ahmad and Kanzi, Sara and Sakall{\i}, {\.I}zzet",
    title = "{Greybody radiation of scalar and Dirac perturbations of NUT black holes}",
    doi = "10.1140/epjp/s13360-021-02227-9",
    journal = "Eur. Phys. J. Plus",
    volume = "137",
    number = "1",
    pages = "94",
    year = "2022",
    url = "https://doi.org/10.1140/epjp/s13360-021-02227-9"
}

@article{AlBadawi2020b,
    author = "Al-Badawi, A. and Sakall{\i}, {\.I}. and Kanzi, S.",
    title = "{Solution of Dirac equation and greybody radiation around a regular Bardeen black hole surrounded by quintessence}",
    journal = "Annals Phys.",
    volume = "412",
    pages = "168026",
    year = "2020",
    url = "https://doi.org/10.1016/j.aop.2019.168026"
}

@article{AlBadawi2020c,
    author = "Al-Badawi, A. and Kanzi, S. and Sakall{\i}, {\.I}.",
    title = "{Greybody factor and Hawking radiation for a Schwarzschild black hole surrounded by quintessence}",
    journal = "Eur. Phys. J. Plus",
    volume = "135",
    pages = "219",
    year = "2020",
    url = "https://doi.org/10.1140/epjp/s13360-020-00245-7"
}

@article{Kanzi2020,
    author = "Kanzi, S. and Mazharimousavi, S. H. and Sakall{\i}, {\.I}.",
    title = "{Greybody factors of black holes in dRGT massive gravity coupled with nonlinear electrodynamics}",
    journal = "Annals Phys.",
    volume = "422",
    pages = "168301",
    year = "2020",
    url = "https://doi.org/10.1016/j.aop.2020.168301"
}

@article{Kanzi2021,
    author = "Kanzi, S. and Sakall{\i}, {\.I}.",
    title = "{Greybody radiation and quasinormal modes of Kerr-like black hole in Bumblebee gravity model}",
    journal = "Eur. Phys. J. C",
    volume = "81",
    pages = "501",
    year = "2021",
    url = "https://doi.org/10.1140/epjc/s10052-021-09299-y"
}

@article{Sakalli2022c,
    author = "Sakall{\i}, {\.I}. and Kanzi, S.",
    title = "{Physical properties of brane-world black hole solutions via a confining potential}",
    journal = "Annals Phys.",
    volume = "439",
    pages = "168803",
    year = "2022",
    url = "https://doi.org/10.1016/j.aop.2022.168803"
}

@article{Gogoi2024,
    author = "Gogoi, Dhruba Jyoti and Heidari, N. and Kriz, J. and Hassanabadi, H.",
    title = "{Quasinormal modes and greybody factors of black holes in modified gravity}",
    journal = "Fortsch. Phys.",
    volume = "72",
    pages = "2300245",
    year = "2024",
    url = "https://doi.org/10.1002/prop.202300245"
}

@misc{Hosseinifar2024,
    author = "Hosseinifar, F. and Ara{\'u}jo Filho, A. A. and Zhang, M. Y. and Chen, H. and Hassanabadi, H.",
    title  ={Shadows, greybody factors, emission rate, topological charge, and phase transitions for a charged black hole with a Kalb-Ramond field background},
    eprint = "2407.07017",
    archivePrefix = "arXiv",
    primaryClass = "gr-qc",
    year = "2024",
    doi   ={https://doi.org/10.48550/arXiv.2407.07017}
}

@article{Sekhmani2025,
    author = "Sekhmani, Y. and Gogoi, D. J. and Maurya, S. K. and Boshkayev, K. and Jasim, M. K.",
    title = "{Effects of modified Chaplygin gas and quintessence on black hole field propagation}",
    journal = "JHEAp",
    volume = "45",
    pages = "200",
    year = "2025",
    url = "https://doi.org/10.1016/j.jheap.2024.12.012"
}

@article{Sakalli2020,
    author = "Sakall{\i}, {\.I}. and Kanzi, S.",
    title = "{Topical review on greybody factors and quasinormal modes in various theories}",
    journal = "Turk. J. Phys.",
    volume = "46",
    number = "2",
    pages = "51",
    year = "2022",
    url = "https://doi.org/10.55730/1300-0101.2691"
}

@article{Skyrme1961a,
    author = "Skyrme, T. H. R.",
    title = "{A non-linear field theory}",
    journal = "Proc. Roy. Soc. Lond. A",
    volume = "260",
    pages = "127--138",
    year = "1961",
    url = "https://doi.org/10.1098/rspa.1961.0018"
}

@article{Skyrme1961b,
    author = "Skyrme, T. H. R.",
    title = "{Particle states of a quantized meson field}",
    journal = "Proc. Roy. Soc. Lond. A",
    volume = "262",
    pages = "237--245",
    year = "1961",
    url = "https://doi.org/10.1098/rspa.1961.0115"
}

@article{Skyrme1962,
    author = "Skyrme, T. H. R.",
    title = "{A unified field theory of mesons and baryons}",
    journal = "Nucl. Phys.",
    volume = "31",
    pages = "556--569",
    year = "1962",
    url = "https://doi.org/10.1016/0029-5582(62)90775-7"
}

@article{Canfora2013a,
    author = "Canfora, Fabrizio and Maeda, Hideki",
    title = "{Hedgehog ansatz and its generalization for self-gravitating Skyrmions}",
    journal = "Phys. Rev. D",
    volume = "87",
    pages = "084049",
    year = "2013",
    url = "https://doi.org/10.1103/PhysRevD.87.084049"
}

@article{Droz1991,
    author = "Droz, S. and Heusler, M. and Straumann, N.",
    title = "{New black hole solutions with hair}",
    journal = "Phys. Lett. B",
    volume = "268",
    pages = "371--376",
    year = "1991",
    url = "https://doi.org/10.1016/0370-2693(91)91592-J"
}

@article{Canfora2013b,
    author = "Canfora, Fabrizio and Correa, Francisco and Giacomini, Alex and Oliva, Julio",
    title = "{Exact meron Black Holes in four dimensional SU(2) Einstein-Yang-Mills theory}",
    journal = "Phys. Lett. B",
    volume = "722",
    pages = "364--371",
    year = "2013",
    url = "https://doi.org/10.1016/j.physletb.2013.04.029"
}

@article{AyonBeato2016,
    author = "Ay{\'o}n-Beato, E. and Canfora, F. and Zanelli, J.",
    title = "{Analytic self-gravitating Skyrmions, cosmological bounces and AdS wormholes}",
    journal = "Phys. Lett. B",
    volume = "752",
    pages = "201--205",
    year = "2016",
    url = "https://doi.org/10.1016/j.physletb.2015.11.065"
}

@article{Astorino2018,
    author = "Astorino, Marco and Canfora, Fabrizio and Lagos, Marcela and Vera, Aldo",
    title = "{Black hole and black string solutions with Skyrme hair}",
    journal = "Phys. Rev. D",
    volume = "97",
    pages = "124032",
    year = "2018",
    url = "https://doi.org/10.1103/PhysRevD.97.124032"
}

@article{Canfora2014,
    author = "Canfora, Fabrizio and Correa, Francisco and Zanelli, Jorge",
    title = "{Exact multisoliton solutions in the four dimensional Skyrme model}",
    journal = "Phys. Rev. D",
    volume = "90",
    pages = "085002",
    year = "2014",
    url = "https://doi.org/10.1103/PhysRevD.90.085002"
}

@article{Canfora2018,
    author = "Canfora, F. and Eiroa, E. F. and Sendra, C. M.",
    title = "{Spherically symmetric black holes with Skyrme hair and their shadows}",
    journal = "Eur. Phys. J. C",
    volume = "78",
    pages = "9",
    year = "2018",
    url = "https://doi.org/10.1140/epjc/s10052-017-5476-3"
}

@article{Adkins1983,
    author = "Adkins, Gregory S. and Nappi, Chiara R. and Witten, Edward",
    title = "{Static properties of nucleons in the Skyrme model}",
    journal = "Nucl. Phys. B",
    volume = "228",
    pages = "552--566",
    year = "1983",
    url = "https://doi.org/10.1016/0550-3213(83)90559-X"
}

@article{Barriola1989,
    author = "Barriola, Manuel and Vilenkin, Alexander",
    title = "{Gravitational Field of a Global Monopole}",
    journal = "Phys. Rev. Lett.",
    volume = "63",
    pages = "341--343",
    year = "1989",
    url = "https://doi.org/10.1103/PhysRevLett.63.341"
}

@article{Vishveshwara1970,
    author = "Vishveshwara, C. V.",
    title = "{Scattering of gravitational radiation by a Schwarzschild black hole}",
    journal = "Nature",
    volume = "227",
    pages = "936--938",
    year = "1970",
    url = "https://doi.org/10.1038/227936a0"
}

@article{Ferrari1984,
    author = "Ferrari, Valeria and Mashhoon, Bahram",
    title = "{New approach to the quasinormal modes of a black hole}",
    journal = "Phys. Rev. D",
    volume = "30",
    pages = "295--304",
    year = "1984",
    url = "https://doi.org/10.1103/PhysRevD.30.295"
}

@article{Zhang2020,
    author = "Zhang, Cheng-Yong and Zhang, Sheng-Jie and Li, Peng-Cheng and Guo, Minyong",
    title = "{Superradiance and stability of the regularized 4D charged Einstein-Gauss-Bonnet black hole}",
    journal = "JHEP",
    volume = "08",
    pages = "105",
    year = "2020",
    url = "https://doi.org/10.1007/JHEP08(2020)105"
}

@book{ref70,
    author = "Wheeler, John A.",
    title = "{Geometrodynamics}",
    publisher = "Academic Press",
    address = "New York",
    year = "1973"
}

@book{ref71,
    author = "Ruffini, A. R.",
    title = "{Black Holes: Les Astres Occlus}",
    publisher = "Gordon and Breach Science Publishers",
    address = "New York",
    year = "1973"
}

@article{Unruh1973,
    author = "Unruh, W. G.",
    title = "{Separability of the neutrino equations in a Kerr background}",
    journal = "Phys. Rev. Lett.",
    volume = "31",
    pages = "1265--1267",
    year = "1973",
    url = "https://doi.org/10.1103/PhysRevLett.31.1265"
}

@article{Chandrasekhar1976,
    author = "Chandrasekhar, S.",
    title = "{The solution of Dirac's equation in Kerr geometry}",
    journal = "Proc. Roy. Soc. Lond. A",
    volume = "349",
    pages = "571--575",
    year = "1976",
    url = "https://doi.org/10.1098/rspa.1976.0090"
}

@article{SS2,
    author = "Schutz, Bernard F. and Will, Clifford M.",
    title = "{Black hole normal modes: a semianalytic approach}",
    journal = "Astrophys. J. Lett.",
    volume = "291",
    pages = "L33--L36",
    year = "1985",
    url = "https://doi.org/10.1086/184453"
}

@inproceedings{SS3,
    author = "Mashhoon, B.",
    title = "{Stability of charged rotating black holes in the eikonal approximation}",
    booktitle = "{Proc. 3rd Marcel Grossmann Meeting on General Relativity}",
    editor = "Ning, H.",
    publisher = "North-Holland",
    address = "Amsterdam",
    pages = "599--608",
    year = "1983"
}

@article{SS4,
    author = "Blome, H.-J. and Mashhoon, B.",
    title = "{Quasi-normal oscillations of a Schwarzschild black hole}",
    journal = "Phys. Lett. A",
    volume = "110",
    pages = "231--234",
    year = "1984",
    url = "https://doi.org/10.1016/0375-9601(84)90058-7"
}

@article{SS5,
    author = "Liu, H. and Mashhoon, B.",
    title = "{On the spectrum of oscillations of a Schwarzschild black hole}",
    journal = "Class. Quant. Grav.",
    volume = "13",
    pages = "233--252",
    year = "1996",
    url = "https://doi.org/10.1088/0264-9381/13/2/009"
}

@article{SS6,
    author = "Iyer, Sai and Will, Clifford M.",
    title = "{Black Hole Normal Modes: A WKB Approach. 1. Foundations and Application of a Higher Order WKB Analysis of Potential Barrier Scattering}",
    journal = "Phys. Rev. D",
    volume = "35",
    pages = "3621",
    year = "1987",
    url = "https://doi.org/10.1103/PhysRevD.35.3621"
}

@article{SS7,
    author = "Konoplya, R. A.",
    title = "{Quasinormal behavior of the D-dimensional Schwarzschild black hole and higher order WKB approach}",
    journal = "Phys. Rev. D",
    volume = "68",
    pages = "024018",
    year = "2003",
    url = "https://doi.org/10.1103/PhysRevD.68.024018"
}

@article{SS8,
    author = "Konoplya, R. A.",
    title = "{Gravitational quasinormal radiation of higher dimensional black holes}",
    journal = "J. Phys. Stud.",
    volume = "8",
    pages = "93--100",
    year = "2004"
}

@article{SS9,
    author = "Matyjasek, Jerzy and Opala, Michal",
    title = "{Quasinormal modes of black holes: The improved semianalytic approach}",
    journal = "Phys. Rev. D",
    volume = "96",
    pages = "024011",
    year = "2017",
    url = "https://doi.org/10.1103/PhysRevD.96.024011"
}

@article{SS10,
    author = "Konoplya, R. A. and Zhidenko, A. and Zinhailo, A. F.",
    title = "{Higher order WKB formula for quasinormal modes and grey-body factors: recipes for quick and accurate calculations}",
    journal = "Class. Quant. Grav.",
    volume = "36",
    pages = "155002",
    year = "2019",
    url = "https://doi.org/10.1088/1361-6382/ab2e25"
}

@article{fb1,
    author = "Schutz, Bernard F. and Will, Clifford M.",
    title = "{Black hole normal modes: a semianalytic approach}",
    journal = "Astrophys. J. Lett.",
    volume = "291",
    pages = "L33--L36",
    year = "1985",
    url = "https://doi.org/10.1086/184453"
}

@article{iyer2,
    author = "Iyer, Sai and Will, Clifford M.",
    title = "{Black Hole Normal Modes: A WKB Approach. 1.}",
    journal = "Phys. Rev. D",
    volume = "35",
    pages = "3621",
    year = "1987",
    url = "https://doi.org/10.1103/PhysRevD.35.3621"
}

@article{Cardoso2009,
    author = "Cardoso, Vitor and Miranda, Alex S. and Berti, Emanuele and Witek, Helvi and Zanchin, Vilson T.",
    title = "{Geodesic stability, Lyapunov exponents and quasinormal modes}",
    journal = "Phys. Rev. D",
    volume = "79",
    pages = "064016",
    year = "2009",
    url = "https://doi.org/10.1103/PhysRevD.79.064016"
}

@article{CB2006,
    author = "Boehmer, Christian G. and Harko, Tiberiu",
    title = "{Can dark matter be a Bose-Einstein condensate?}",
    journal = "Class. Quant. Grav.",
    volume = "23",
    pages = "6479--6491",
    year = "2006",
    url = "https://doi.org/10.1088/0264-9381/23/22/017"
}

@article{SF2005,
    author = "Fernando, Sharmanthie",
    title = "{Greybody factors of charged dilaton black holes in 2+1 dimensions}",
    journal = "Gen. Rel. Grav.",
    volume = "37",
    pages = "585--604",
    year = "2005",
    url = "https://doi.org/10.1007/s10714-005-0049-4"
}

@article{Konoplya2023,
    author = "Konoplya, R. A.",
    title = "{The sound of the event horizon}",
    journal = "Int. J. Mod. Phys. D",
    volume = "32",
    number = "14",
    pages = "2342014",
    year = "2023",
    url = "https://doi.org/10.1142/S0218271823420142"
}

@article{LIGO2,
    author = "Abbott, B. P. and others",
    collaboration = "LIGO Scientific, Virgo",
    title = "{Localization and Broadband Follow-up of the Gravitational-wave Transient GW150914}",
    journal = "Astrophys. J. Lett.",
    volume = "826",
    pages = "L13",
    year = "2016",
    url = "https://doi.org/10.3847/2041-8205/826/1/L13"
}

@article{LIGO4,
    author = "Abbott, R. and others",
    collaboration = "LIGO Scientific, Virgo",
    title = "{GW190412: Observation of a Binary-Black-Hole Coalescence with Asymmetric Masses}",
    journal = "Phys. Rev. D",
    volume = "102",
    pages = "043015",
    year = "2020",
    url = "https://doi.org/10.1103/PhysRevD.102.043015"
}

@article{Deng2026,
    author = "Liang, Jie and Liu, Dong and Long, Zheng-Wen",
    title = "{Quasinormal modes and greybody factors of black holes corrected by nonlinear electrodynamics}",
    journal = "Eur. Phys. J. C",
    volume = "86",
    pages = "17",
    year = "2026",
    url = "https://doi.org/10.1140/epjc/s10052-025-15245-z"
}

@article{Singh2024,
    author = "Sekhmani, Y. and Gogoi, D. J. and Maurya, S. K. and Boshkayev, K. and Jasim, M. K.",
    title = "{Quasinormal modes and greybody bounds of black holes endowed with modified Chaplygin gas}",
    journal = "J. High Energy Astrophys.",
    volume = "45",
    pages = "200--214",
    year = "2025",
    url = "https://doi.org/10.1016/j.jheap.2024.11.008"
}

@article{Bolokhov2024,
    author = "Bolokhov, S. V.",
    title = "{Long-lived quasinormal modes and overtones' behavior of holonomy-corrected black holes}",
    journal = "Phys. Rev. D",
    volume = "110",
    pages = "024010",
    year = "2024",
    url = "https://doi.org/10.1103/PhysRevD.110.024010"
}

\end{document}